\documentclass[fleqn,usenatbib]{mnras}

\usepackage{newtxtext,newtxmath}

\usepackage[T1]{fontenc}

\DeclareRobustCommand{\VAN}[3]{#2}
\let\VANthebibliography\thebibliography
\def\thebibliography{\DeclareRobustCommand{\VAN}[3]{##3}\VANthebibliography}

\usepackage{graphicx}	
\usepackage{amsmath}	
\usepackage{xcolor}
\usepackage{comment}
\usepackage{pdflscape}
\usepackage{tablefootnote}
\usepackage{natbib}
\usepackage{threeparttable}

\title[New DIBs found in residuals of stellar spectra]{The GALAH survey: New diffuse interstellar bands found in residuals of 872,000 stellar spectra}

\author[R. Vogrinčič et al.]{Rok Vogrinčič$^{1}$\thanks{E-mail: rok.vogrincic@fmf.uni-lj.si}, Janez Kos$^{1}$, Tomaž Zwitter$^{1}$, Gregor Traven$^{1}$, Kevin L. Beeson$^{1}$, Klemen Čotar$^{8,1}$,
\newauthor
Ulisse Munari$^{2}$, Sven Buder$^{6,5}$, Sarah L. Martell$^{7,5}$, Geraint F. Lewis$^{3}$, Gayandhi M De Silva$^{4,5}$, 
\newauthor
Michael R. Hayden$^{3,5}$, Joss Bland-Hawthorn$^{3,5}$, Valentina D'Orazi$^{9,10,11}$
\\
$^{1}$Faculty of Mathematics and Physics, University of Ljubljana, Jadranska 19, 1000 Ljubljana, Slovenia\\
$^{2}$INAF Astronomical Observatory of Padova, I-36012 Asiago (VI), Italy
\\
$^{3}$Sydney Institute for Astronomy, School of Physics, A28, The University of Sydney, NSW 2006, Australia\\
$^{4}$Australian Astronomical Optics, Macquarie University, 105 Delhi Rd, North Ryde, NSW 2113, Australia\\
$^{5}$Centre of Excellence for All-Sky Astrophysics in Three Dimensions (ASTRO 3D), Australia\\
$^{6}$Research School of Astronomy $\&$ Astrophysics, Australian National University, Canberra ACT 2611, Australia\\
$^{7}$School of Physics, University of New South Wales, Sydney NSW 2052, Australia\\
$^{8}$Flai d.o.o., Bravničarjeva ulica 13, 1000 Ljubljana, Slovenia\\
$^{9}$Dipartimento di Fisica, Università degli Studi di Roma Tor Vergata, via della Ricerca Scientifica 1, 00133, Roma, Italy\\
$^{10}$INAF Osservatorio Astronomico di Padova, Vicolo dell’Osservatorio 5, 35122, Padova, Italy\\
$^{11}$Department of Physics and Astronomy, Monash University, Clayton, VIC 3800, Australia\\
}

\date{Accepted XXX. Received YYY; in original form ZZZ}

\pubyear{2022}

\begin{document}

\defcitealias{Jenniskens1994}{JD94}
\defcitealias{HobbsYork2009}{HY09}
\defcitealias{GalazutdinovMusaev2000}{GM00}
\defcitealias{Tuairisg2000}{TC00}
\defcitealias{Weselak2000}{WS00}
\defcitealias{Fan2019}{FH19}
\defcitealias{Sonnentrucker2018}{SY18}

\label{firstpage}
\pagerange{\pageref{firstpage}--\pageref{lastpage}}
\maketitle

\begin{abstract}

We use more than 872,000 mid-to-high resolution (R $\sim$ 20,000) spectra of stars from the GALAH survey to discern the spectra of diffuse interstellar bands (DIBs). We use four windows with the wavelength range from 4718 to 4903, 5649 to 5873, 6481 to 6739, and 7590 to 7890 \AA, giving a total coverage of 967 \AA. We produce $\sim$400,000 spectra of interstellar medium (ISM) absorption features and correct them for radial velocities of the DIB clouds. Ultimately, we combine the 33,115 best ISM spectra into six reddening bins with a range of $0.1 \,\mathrm{mag} < E\mathrm{(B-V)} < 0.7\, \mathrm{mag}$. A total of 183 absorption features in these spectra qualify as DIBs, their fitted model parameters are summarized in a detailed catalogue. From these, 64 are not reported in the literature, among these 17 are certain, 14 are probable and 33 are possible. We find that the broad DIBs can be fitted with a multitude of narrower DIBs. Finally, we create a synthetic DIB spectrum at unit reddening which should allow us to narrow down the possible carriers of DIBs and explore the composition of the ISM and ultimately better model dust and star formation as well as to correct Galactic and extragalactic observations. The majority of certain DIBs show a significant excess of equivalent width when compared to reddening. We explain this with observed lines of sight penetrating more uniform DIB clouds compared to clumpy dust clouds. 
\end{abstract}

\begin{keywords}
methods: data analysis – Astronomical data bases: catalogues - Astronomical data bases: surveys – ISM: dust, extinction
\end{keywords}



\section{Introduction}

Diffuse interstellar bands are absorption lines at visual and near infrared wavelengths observed in spectra of stars reddened by dust. DIBs have an origin in the ISM and most of their carriers are still not identified \citep{Herbig1995}. The identification of DIBs therefore remains one of the longest standing problems in astronomical spectroscopy \citep{Herbig1995, Sarre2006}.

There are over 400 known DIBs, which are resolved in mid-to-high resolution spectra and they show a wide range of intensities and widths \citep{Sarre2006}, from narrowest having a full width at half-maximum ($FWHM$) of about 0.5 \AA, to broadest having a $FWHM$ greater than 10 \AA. DIBs are commonly designated by their rest wavelength rounded to the nearest \AA; e.g., DIB at 5780.4 \AA~is called DIB 5780. All DIBs observed in this survey are optically thin, i.e. their equivalent widths ($EW$s) are linearly proportional to the column densities of the absorbing material. Their strengths are expressed as $EW$s instead of column densities, because their oscillator strengths are unknown \citep{Kos2017}. The measured strengths of DIBs vary depending on the line of sight (LoS) where they are observed and are correlated with the amount of interstellar reddening \citep{Krelowski2019, Zasowski2015, Lan2015, KosZwitter2013, Friedman2011, Megier2005}

For decades, DIBs were observed in high signal-to-noise ratio ($SNR$) spectra of hot stars where weak DIBs are unlikely to be blended with stellar lines \citep{Cami1997, Friedman2011, Puspitarini2013}. Surveys of hot stars have included a few thousand stars at most \citep{VanLoon2014}, and in some cases, when investigating weak DIBs, only around a hundred \citep[e.g.][]{Friedman2011}. In such surveys a small number of hot stars in the Galaxy is a strong limitation of the number of LoSs they can observe. With the rise of large spectroscopic sky surveys in recent years, the number of observed LoSs increased tremendously, thus motivating us to detect DIBs in a much wider range of stellar types, as we are not limited only to hot stars.

The idea to observe DIBs in the GALAH survey \citep{Buder2021} was first introduced as part of the ancillary science motivation in \citet{DeSilva2015}. We show that a large number of mid-to-high resolution spectra of cool stars in the GALAH survey can indeed be used as a tool to detect DIBs with amplitudes of less than 1$\%$ below the continuum. This is possible because of the large number of observed LoSs, with $E\mathrm{(B-V)} \geq 0.1$ mag. We use an efficient method for detecting the DIBs in spectra of cool stars, described in detail by \citet{Kos2013}, where a stellar template spectrum is constructed from observed spectra that lack interstellar features and that match the stellar component of the observed spectrum in question. The observed spectrum and the template spectrum are divided and their quotient presents an absorption spectrum of the interstellar medium (ISM) where DIBs can be detected. \citet{YuanLiu2012} used a similar method, where they identified strong DIBs in the SDSS spectra of a sample of $\sim$2000 stars with $E\mathrm{(B-V)} \sim 0.2-1.0$ mag and measured their strengths and radial velocities.

The goal of our survey is to expand the list of known DIBs and to verify the existing parameters of the DIBs reported in the literature. Modern catalogues of DIBs were built mainly by means of observing hot and strongly reddened stars.  Jenniskens et al. \citepalias[1994; hereafter][]{Jenniskens1994} detected 229 DIBs in the wavelength range from 3800 \AA~to 8680 \AA~with 0.3 \AA~resolution in spectra of four reddened hot stars, HD 30614 ($E\mathrm{(B-V)} = 0.3$ mag), HD 21389 ($E\mathrm{(B-V)} = 0.54$ mag), HD 190603 ($E\mathrm{(B-V)} = 0.72$ mag) and HD 183143 ($E\mathrm{(B-V)} = 1.28$ mag). Galazutdinov et al. \citepalias[2000; hereafter][]{GalazutdinovMusaev2000} detected 271 DIBs, of which more than 100 were newly discovered, in the wavelength range from 4460 \AA~and 8800 \AA~based on six spectra of reddened stars with a high spectral resolution of $\sim$45,000. Similarly, Tuairisg et al. \citepalias[2000; hereafter][]{Tuairisg2000} detected 226 DIBs, of which 60 were newly discovered. The survey was done in the wavelength range between 3906 \AA~and 6812 \AA~toward three very reddened and five unreddened stars. Weselak et al. \citepalias[2000; hereafter][]{Weselak2000} found 213 DIBs in medium reddened (0.1 - 0.5 mag) early type stars in a wavelength range from 5650 \AA~to 6865 \AA. Hobbs et al. \citepalias[2009; hereafter][]{HobbsYork2009} presented an extensive catalogue of 414 DIBs (135 DIBs were new, mostly weak and narrow), measured between 3900 \AA~and 8100 \AA~in a spectrum with a resolving power of $\sim$38,000 of the star HD 183143 which has the reddening of $E\mathrm{(B-V)} = 1.27$ mag. Fan et al. \citepalias[2019; hereafter][]{Fan2019} detected 559 DIBs of which 22 are new DIBs, and adjusted the boundaries of 27 sets of DIBs observed in 25 medium to highly reddened LoSs towards hot (O, B, A type) stars. A comprehensive study was made in the wavelength range between 4000 \AA~and~9000 \AA, with median $SNR \sim$ 1300 at 6400 \AA. Sonnentrucker et al. \citepalias[2018; hereafter][]{Sonnentrucker2018} observed and investigated 22 broad DIBs ($FWHM \geq$ 6 \AA) in low-resolution stellar spectra measured between 3600 \AA~and 9000 \AA~in 21 LoSs in a reddening range of three magnitudes, towards hot stars (O7 through A3). Nine out of 22 features showed increasing strength ($EW$) with reddening, 11 features were categorized as possible DIBs as they were affected by blending with stellar and telluric absorption. Two features were rejected as DIBs. 
Although these surveys provide a large number of DIBs, they are confined only to a small sample of highly reddened stars. Observing a large number of LoSs with a varying amount of interstellar reddening gives us a better insight into the relationship between the amount of reddening and the strengths of the DIBs. This in turn provides more meaningful estimates of DIB parameters. 

This paper is ordered as follows. Section \ref{Data Analysis} presents the method we use to extract DIBs from spectra of all but predominantly cool stars in the GALAH survey. Section \ref{Results} contains the results and discussion. We show the averaged ISM absorption spectra in three wavelength bands ranging from 4700 \AA~to 6800 \AA, with average reddening $E\mathrm{(B-V)}$ of 0.15, 0.25, 0.35, 0.45, 0.55 and 0.65 magnitude. We present a catalogue of identified DIBs with measured parameters (central wavelength, $FWHM$, asymmetry index, slope and intercept of the regression line in the relation of equivalent width with respect to the reddening) and their cross-reference with DIBs from other catalogues. Additionally, we show the synthetic DIB spectrum scaled to reddening $E\mathrm{(B-V)} =$ 1 mag that was built using the measured DIB parameters. Section \ref{conclusion} contains conclusions. Appendix \ref{sec:appendix_A} contains figures which demonstrate our method of finding the DIB candidates used for fitting the averaged ISM spectra. Appendix \ref{sec:appendix_B} contains a complete catalogue of parameters of 183 identified DIBs.  

\section{Data and analysis} \label{Data Analysis}

We use spectra from data release 4 of the GALAH survey (GALAH collaboration, in prep.), which includes 872,228 observed stars, each star corresponding to one LoS. GALAH \citep{Buder2021} covers about a half of the sky and it uses the HERMES multi-object high-resolution spectrograph \citep{HERMES2015} on the 3.9 m Anglo-Australian Telescope. The instrument produces spectra with $SNR \sim$ 100 per resolution element ($\lambda$ / $\delta \lambda \lesssim$ 28,000) at V = 14 in a one to one and a half hour long exposure. Its four spectral bands cover about 1000 \AA~in total between 4718 and 7890 \AA~\citep{Zucker2012, Buder2021}. The majority of observed stars in GALAH are distributed within 4 kpc from the Sun, at galactic latitudes $\mathrm{|b|} > 10$ deg. Relatively few stars are observed close to the Galactic plane, and they are mainly in the direction of the Galactic Centre \citep{Buder2021}.

Our goal is to detect DIBs in as many LoSs as possible without relying on hot stars which are scarce. Of the 872,228 available stars in the GALAH survey, only 10,466 (1.2$\%$) have colour index (BP - RP), obtained from Gaia EDR3 archive, \citep{GaiaEDR3_2021} lower than 0.4 mag, which corresponds to hot stars (O, B and A type stars), where DIBs can be easily distinguished from stellar lines. Cooler stars have spectra populated by more stellar absorption lines, thus DIBs can be blended with them and cannot be studied directly. In order to detect DIBs in the spectra of the stars of later spectral types, we follow the procedure described in Section \ref{Data reduction}. 

\begin{figure}
\begin{center}
\includegraphics[scale=1]{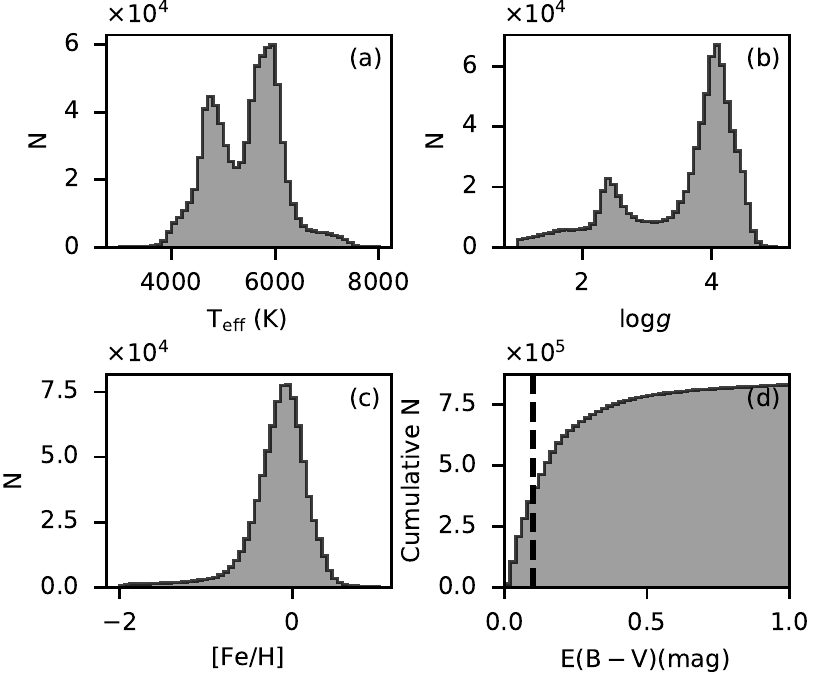}
\caption[f1]{A distribution of \textbf{(a)} effective temperature, \textbf{(b)} surface gravity, \textbf{(c)} metallicity and \textbf{(d)} reddening of stars in the GALAH DR4 survey. The vertical dashed line at $E\mathrm{(B-V)} = 0.1$ mag denotes a border above which 60$\%$ of stars lie.}
\label{fig:Figure 1}
\end{center}
\end{figure}

A vast majority of stars in the GALAH survey have temperatures between 4000 K and 7500 K, with a median of 5550 K. In Fig. \ref{fig:Figure 1} we find two groups of stars, one with a peak at around 4700 K, the other at around 6000 K. A similar pattern is seen in the surface gravity domain, where the first peak in $\mathrm{\log \it{g}}$ at around 2.5 corresponds to cool giants which represent about 26$\%$ of all stars in the GALAH survey. The second peak in $\mathrm{\log \it{g}}$ at around 4 corresponds to main sequence stars with temperatures of about 6000 K. Main sequence stars represent about 69$\%$ of all stars in the GALAH survey. The remaining 5$\%$ are peculiar stars. Metallicity of stars in the GALAH survey shows only one peak at around $-0.1$, which means that the stars from the Galactic thin and thick disk in a volume centered on the Sun are the most numerous. Stars also have a substantial range of interstellar dust reddening, i.e. about 60$\%$ of all stars have reddening $E\mathrm{(B-V)} > 0.1$ mag, as seen in a cumulative distribution of reddening obtained from the SFD catalogue, \citep{SFD2011} in Fig. \ref{fig:Figure 1}. Note that most of the targets are located away from the galactic plane, so that their reddening is similar to the cumulative value. Altogether, stars in the GALAH survey prove to be favourable targets in the search of DIBs, as they have a decent amount of reddening needed to detect them.

\subsection{Data reduction} \label{Data reduction}
\subsubsection{Continuum normalization}  \label{Continuum}

In order to detect DIBs in spectra of cool stars, one must first ensure that the spectra used in the analysis are normalized with good precision. The spectra from data release 4 of the GALAH survey, (the GALAH collaboration, in prep.) are non-parametrically normalized, meaning that the normalization is done without the prior knowledge of the stellar parameters. Because the majority of DIBs have amplitudes of the order of 0.1$\%$ below the continuum, it is required that the re-normalization is done parametrically. This is especially important when combining the spectra to increase the $SNR$ in order to reveal weak DIBs.

For the purpose of re-normalization we use the sigma-clipping procedure which removes outliers (absorption and emission lines in the spectra) and allows for a more robust estimate of the continuum. To do that, we need to obtain the optimal sigma-clipping parameters ($\sigma_{\mathrm{lo}}$, $\sigma_{\mathrm{hi}}$) as a function of stellar parameters and $SNR$. These parameters are derived when replicating the known continuum of a synthetic spectrum with added noise.

Our re-normalization procedure starts with the creation of a 4-D grid of synthetic spectra from the \citet{Kurucz2003} library, with $T_{\mathrm{eff}}$ (effective temperature, ranging from 3,500 K to 10,000 K, with a step size of 250 K), $\mathrm{\log \it{g}}$ (surface gravity, ranging from 0 to 5 dex, with a step size of 0.5 dex), $\mathrm{[Fe/H]}$ (metallicity, ranging from -2.5 dex to 0.5 dex, with a step size of 0.5 dex) and $SNR$ (ranging from 10 to either 120, 190 or 250 for blue, green, and red/IR spectral arms respectively, with a step size of 10). Here, $SNR$ is not a parameter of a synthetic spectrum but a parameter which simulates synthetic spectrum with added noise. Note also that some combinations of the above-mentioned parameters of synthetic spectra are not available in the \citet{Kurucz2003} library, therefore some nodes in the 4-D grid remain empty. Moreover, synthetic spectra from \citet{Kurucz2003} are used only to explore and fit the sigma-clipping parameters.

\begin{figure}
\begin{center}
\includegraphics[scale=1]{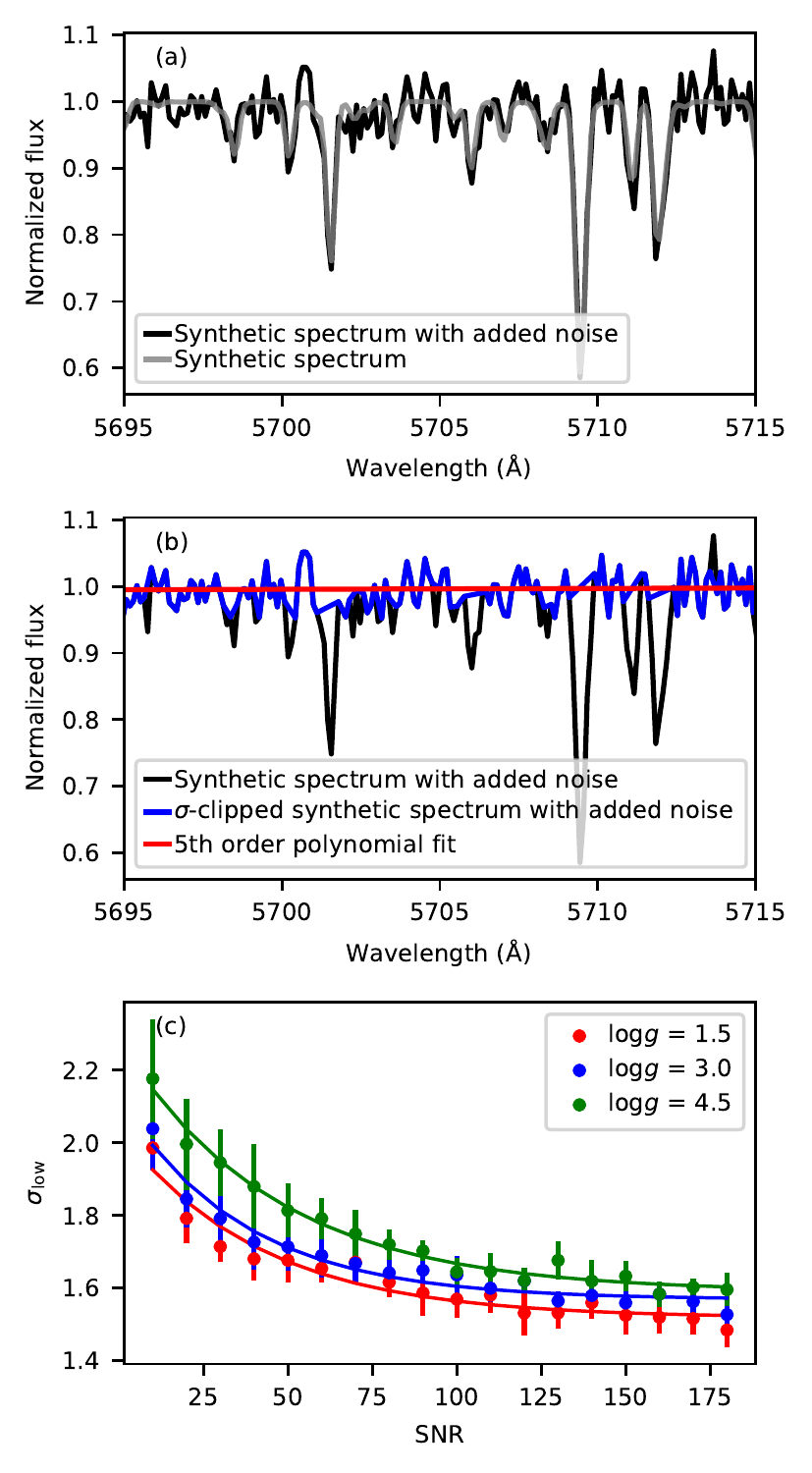}
\caption[f1]{\textbf{(a)} An example of a synthetic spectrum (gray) and a synthetic spectrum with Poissonian noise added (black) of a star with $T_{\mathrm{eff}}$ = 4750 K, $\mathrm{\log \it{g}}$ = 2.5 and $\mathrm{[Fe/H]}$ = -1.5, in green spectral arm. \textbf{(b)} A synthetic spectrum with Poissonian noise added (black), a sigma-clipped synthetic spectrum with Poissonian noise added (blue) and a 5$^{\mathrm{th}}$ order polynomial fit to a sigma-clipped spectrum (red). \textbf{(c)} A trend of $\sigma_{\mathrm{lo}}$ with respect to $SNR$, for a star with a constant temperature $T_{\mathrm{eff}}$ = 4750 K and metalicity $\mathrm{[Fe/H]} = -1.5$, with a varying surface gravity  $\mathrm{\log \it{g}}$.}
\label{fig:Figure 2A}
\end{center}
\end{figure}

Fig. \ref{fig:Figure 2A} illustrates the sigma-clipping procedure used in our analysis. Fig. \ref{fig:Figure 2A} (a) shows an example of a synthetic spectrum (gray line) of a star with $T_{\mathrm{eff}}$ = 4750 K, $\mathrm{\log \it{g}}$ = 2.5 and $\mathrm{[Fe/H]}$ = -1.5, in the green spectral arm. We simulate $SNR$ by adding Poissonian noise to the spectrum (black line). Fig. \ref{fig:Figure 2A} (b) illustrates the sigma-clipping algorithm performed on the noisy spectrum (black line) and fitting of the sigma-clipped synthetic spectrum (blue line) with a 5$^{\mathrm{th}}$ order least squares polynomial (red line) over the whole spectral arm. At each node of the grid we produce 10 different variants of the Poissonian noise and we vary $\sigma_{\mathrm{lo}}$ and $\sigma_{\mathrm{hi}}$ so that the fitted polynomial is as close to unity as possible. This is done in 12 sigma-clipping iterations for each realization of the Poissonian noise. We then take the  $\sigma_{\mathrm{lo}}$ and $\sigma_{\mathrm{hi}}$ averaged over all Poissonian realizations as the final sigma-clipping parameters. Empty nodes in the 4-D grid are filled with approximate sigma-clipping parameters obtained by interpolating or extrapolating sigma-clipping parameters from adjacent nodes. Fig. \ref{fig:Figure 2A} (c) shows a trend of $\sigma_{\mathrm{lo}}$ with respect to $SNR$, for a star with a constant effective temperature $T_{\mathrm{eff}}$ = 4750 K, metalicity $\mathrm{[Fe/H]}$ = -1.5 and 3 different values of surface gravity $\mathrm{\log \it{g}}$. The values of sigma-clipping parameters change predictably and smoothly, so interpolation and extrapolation are justified. 

In the final step, we use the sigma-clipping parameters to clip the observed spectra and fit the continua. For this we must know the stellar parameters of all observed spectra, which are provided by the reduction pipeline of the data release 4 of the GALAH survey (the GALAH collaboration, in prep.). We take the sigma-clipping parameters from the closest grid point to the observed spectrum. 

\subsubsection{Resolution equalization}
To detect weak DIBs in cool stars effectively, a template spectrum must be constructed from spectra of stars with very similar parameters (see Section \ref{Finding similar spectra} and Section \ref{Template creation}). To ensure that the spectra used to form a template are as similar as possible, they must have very similar resolution. Our spectra vary substantially in resolution (from 20,000 to 28,000) both from object to object and as a function of wavelength, thus it is crucial to equalize resolution for all spectra. The lowest resolution of the spectra in our data is slightly above 20,000. We cannot increase the resolution but we can degrade it to a common value, which in our case is 20,000 for all spectral arms. Such blurring does not widen the observed DIBs substantially but it limits the minimum width of a DIB to 15 km/s.

For each spectrum, a resolution profile (given by a generalized Gaussian function) is provided by the reduction pipeline of the data release 4 of the GALAH survey (the GALAH collaboration, in prep.). To equalize the resolution we re-sample the data (change even sampling to uneven sampling) in such a way that the resolution can be degraded by a convolution with a fixed-width kernel. The data is then re-sampled to the observed pixel spacing to produce a spectrum with a constant resolution profile.

\subsection{DIBs extraction} 

After the reduction procedures described in \ref{Data reduction} all spectra are re-normalized and have equal and constant resolution (R = 20,000). Our next goal is to separate available spectra in two groups: spectra of reddened stars ($E\mathrm{(B-V)} \geq$ 0.1 mag in our case) and spectra of low-reddened stars ($E\mathrm{(B-V)} < 0.1$ mag). From spectra of low-reddened stars we will construct template spectra that lack the ISM absorption lines. Template spectra will be used to remove the stellar component from the spectra of reddened stars. The removal is done only between the spectra of stars that have very similar atmospheric parameters, i. e. similar effective temperature, surface gravity and metalicity. We group spectra according to atmospheric parameters of stars listed in the GALAH survey (data release 4). For a spectrum of a reddened star we average its low-reddened counterparts into a template and divide it from the reddened one. The resulting spectrum is called the ISM spectrum, which reveals DIBs that were previously blended with stellar lines in the reddened spectrum. This method is the same as described in \citet{Kos2013}. 

\subsubsection{Target selection} \label{Target selection}

\begin{figure}
\begin{center}
\includegraphics[scale=1]{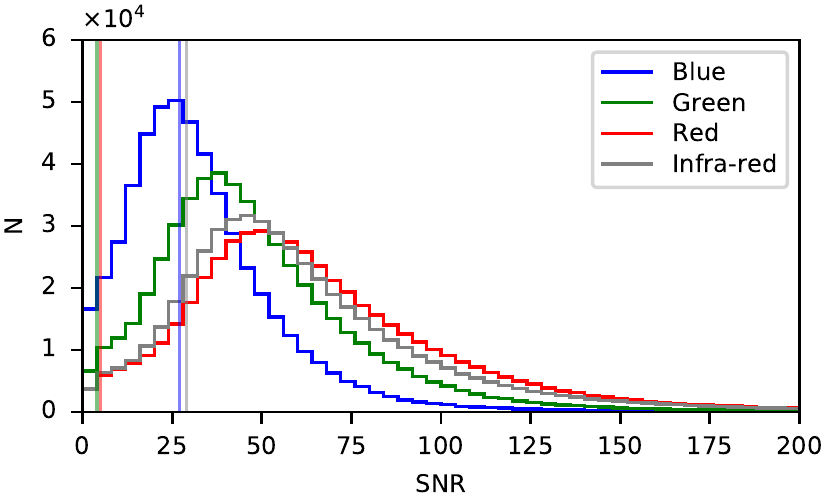}
\caption[f1]{The distribution of $SNR$ of target spectra ($E\mathrm{(B-V)} \geq 0.1$ mag) in the GALAH DR4 survey for the four spectral arms (blue, green, red and infra-red). Vertical lines represent the  adopted lowest useful $SNR$.}
\label{fig:Figure 2}
\end{center}
\end{figure}

The majority of DIBs have amplitudes of the order of 0.1$\%$ below the continuum. Hence, our spectra of reddened stars ($E(\mathrm{B-V}) \geq$ 0.1 mag, from here-on referred to as target spectra), need to have sufficient $SNR$ in order to detect DIBs in them. For that reason we estimate the lowest useful $SNR$ for each spectral arm (see Table \ref{tab:primer_tabele}). This estimate is obtained so that the strongest DIB can be detected in the spectra with the lowest $E\mathrm{(B-V)} = 0.1$ mag. From the DIB catalogue \citetalias{Jenniskens1994} we take the $FWHM$ and the equivalent width values of DIB 4780, DIB 5780, DIB 6613 and DIB 7721 for blue, green, red and infra-red spectral arm respectively. We calculate the expected amplitude of the DIB at the reddening of $E\mathrm{(B-V)} = 0.1$ mag. To detect these DIBs, their amplitude must be greater than 3 standard deviations ($\sigma$) of the noise of 1000 combined spectra with the lowest useful $SNR$. For blue, green, red and infra-red spectral arms the estimated lowest useful $SNR$s are 27, 4, 5 and 29, respectively. Lowest useful $SNR$ is thus used as a lower limit when selecting suitable targets. The distribution of the $SNR$ of the target spectra is shown in Fig. \ref{fig:Figure 2}. Target spectra with $SNR$ greater than the lowest useful $SNR$ represent from 55$\%$ (blue arm) to 99$\%$ (green and red arm) of all target spectra in the GALAH DR4 survey.

\begin{table*}
    \centering
    \caption{The number of target spectra ($N_{\mathrm{target}}$ ($SNR \geq SNR_{\mathrm{low}}$)) we use in our analysis based on the adopted lowest useful $SNR$, which is denoted by $SNR_{\mathrm{low}}$. Values are accompanied with a fraction of all target spectra ($N_{\mathrm{target}}$ with $E\mathrm{(B-V)} \geq 0.1$ mag) in the GALAH DR4 survey. $N_{\mathrm{all}}$ presents the number of all spectra per spectral arm. $N_{\mathrm{neighbour}}$ shows the number of nearest neighbour spectra with $E\mathrm{(B-V)} < 0.1$ mag that were used in order to build templates. The wavelength range covered by each arm of the HERMES spectrograph is denoted by $\lambda_{\mathrm{min}}$ and $\lambda_{\mathrm{max}}$.}
    \label{tab:primer_tabele}
    \begin{tabular}{l c c c c c c l}
    \hline
        Arm & $\lambda_{\mathrm{min}}$ (\AA) & $\lambda_{\mathrm{max}}$ (\AA) & $SNR_{\mathrm{low}}$ & $N_{\mathrm{all}}$ & $N_{\mathrm{neighbour}}$ & $N_{\mathrm{target}}$ & $N_{\mathrm{target}}$ ($SNR \geq SNR_{\mathrm{low}}$)\\\hline\hline
         Blue & 4718 & 4903 & 27 & 868,622 & 347,799 & 520,823 & 286,761 (55$\%$)\\
         Green & 5649 & 5873 & 4 & 869,867 & 348,187 & 521,680 & 515,085 (99$\%$)\\
         Red & 6481 & 6739 & 5 & 868,763 & 347,761 & 521,002 & 515,947 (99$\%$)\\
         Infra-red & 7590 & 7890 & 29 & 853,179 & 343,059 & 510,120 & 437,277 (86$\%$)\\\hline
    \end{tabular}
\end{table*}

\subsubsection{Finding similar spectra} \label{Finding similar spectra}

The spectra of low-reddened stars ($E\mathrm{(B-V)} <$ 0.1 mag) are used to construct a template spectrum. For every target spectrum we create a sample of nearest neighbour candidates among the spectra of low-reddened stars, based on their effective temperature, surface gravity and metalicity. A star qualifies as a nearest neighbour candidate if its parameters are within $T_{\mathrm{eff}}$ $\pm$ 250 K, $\mathrm{\log(g)}$ $\pm$ 0.5 dex and $\mathrm{[Fe/H]}$ $\pm$ 0.5 dex. This step serves as a pre-selection of spectra, so that the process of finding the nearest neighbours is considerably shorter. The sample sizes vary from a few hundred to some tens of thousands of low-reddened spectra.

\subsubsection{Nearest neighbours and template creation}
\label{Template creation}

\begin{figure}
\begin{center}
\includegraphics[scale=.65]{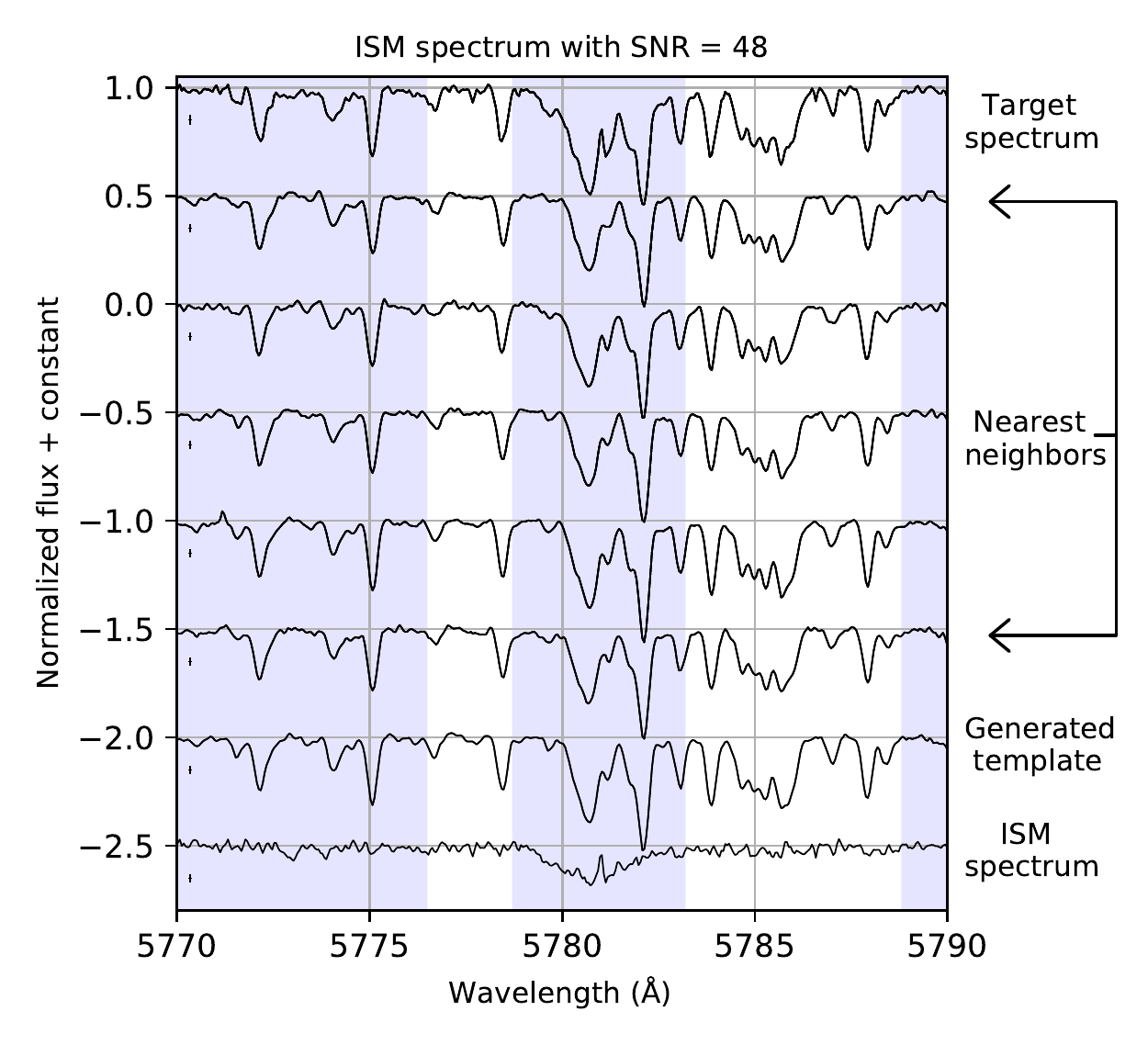}
\caption[f1]{Illustration of the main steps of the procedure to obtain the ISM spectrum. The target spectrum has $SNR$ = 48; parameters of the target are $T_{\mathrm{eff}}$ = 4715 K, $\log g$ = 2.35, and $\mathrm{[Fe/H]}$ = -0.1. From top to bottom the following spectra are shown: the target spectrum (with $E\mathrm{(B-V)}$ = 0.96 mag), which is the spectrum from which we want to extract the DIB. The next 5 spectra are examples of nearest neighbour spectra of low-reddened stars ($E\mathrm{(B-V)}$ < 0.1 mag). Next, 100 such spectra are averaged into a template. The target spectrum is then divided by the template spectrum to get the ISM spectrum on the bottom. Relative flux errors are presented on the bottom left side of each spectrum. A slight dip is noticed in the ISM spectrum, at a wavelength of around 5780 \AA, which is known as DIB 5780. Vertical blue bands indicate regions where strong DIBs and telluric lines are masked, in order to find the most similar spectra unaffected by them.}
\label{fig:Figure 4}
\end{center}
\end{figure}

First, all spectra are corrected for their intrinsic stellar radial velocity ($v_r$) provided by the reduction pipeline of the data release 4 of the GALAH survey (GALAH collaboration, in prep.). Next, we calculate the similarity of the target spectrum and each of its nearest neighbour candidates. The estimator for the similarity of the spectra is the sum of absolute differences of all pixels
\begin{equation}
d(t,n) = \sum_{i} \frac{|t_i - n_i|}{|t_i|+|n_i|} w,
\end{equation}
where $d$ is the calculated distance, $t$ and $n$ are arrays containing normalized fluxes of a target and a neighbour and $w$ is a mask. A target spectrum and its nearest neighbours are masked around strong DIBs, peaks of H$\alpha$ and H$\beta$ absorption and strong telluric lines such as: H$_2$O, O$_2$ and Chappuis ozone absorption bands \citep{Molecfit}. Excluding the aforementioned parts of the spectrum allows an unbiased comparison of a target spectrum and a spectrum of a nearest neighbour candidate. We then take 100 nearest neighbours with greatest similarity and average them into a template spectrum. If a target spectrum has less than 100 neighbours, it is discarded from further analysis as it is deemed too exotic. Each target spectrum is then divided by its appropriate template spectrum. Their quotient is called the ISM spectrum \citep{Kos2013}. The process of obtaining the ISM spectrum is illustrated in Fig. \ref{fig:Figure 4}.

We obtained a large number of ISM spectra. In the blue arm 44$\%$ (228,861) of initially selected targets passed through the aforementioned procedure successfully. Substantially more in the green arm 76$\%$ (398,310), in the red arm 78$\%$ (405,555) and in the infra-red arm 76$\%$ (387,091).

\subsection{Obtaining the DIB parameters}
\subsubsection{Radial velocity correction} \label{RV corr}

Our goal is to combine ISM spectra into an averaged ISM spectrum in which we will then detect DIBs and create a DIB catalogue. To achieve that, we must ensure that each ISM spectrum is corrected for radial velocity of the cloud of matter which absorbs light, thus causing DIBs (henceforth DIB cloud). This part is crucial when searching for very weak and narrow DIBs. Combining all spectra without the radial velocity correction would lead to incorrect estimates of DIB parameters (width, central wavelength and equivalent width) as DIBs would appear weaker and broader.

The procedure of obtaining a radial velocity of a DIB cloud is the following. We do a cross-correlation between small portions (20 \AA~ wide) of the ISM spectrum in question and the synthetic spectrum \citetalias{Jenniskens1994}. Each portion of the spectrum is centered on the wavelength of a strong DIB, i. e. DIB 4780, 5780, 5797, 5850, 6613, 6660, 6699 and 7722. The cross-correlation is performed in the radial velocity frame ranging from -125 km/s to 125 km/s, with a step size of 1 km/s. For each DIB we fit the obtained cross-correlation function with a Gaussian and find the peak correlation coefficient (C) and the radial velocity shift. All spectra with C < 0.4 are discarded. Clearly, not all spectra are suitable for the radial velocity correction but a large number of them are, i. e. 2107 ISM spectra for DIB 4780, 266,498 for DIB 5780, 123,688 for DIB 5797, 11,207 for DIB 5850, 171,507 for DIB 6613, 39,944 for DIB 6660, 17,287 for DIB 6699 and 2014 for DIB 7722. We note that number of ISM spectra in the green and red spectral arms is one to two orders of magnitude larger than in the blue and infra-red ones. DIB 5780 is the strongest and most common DIB in our ISM spectra, therefore its measured radial velocities can be used to shift ISM spectra and to combine them into an averaged ISM spectrum.

\begin{figure}
\begin{center}
\includegraphics[scale=1]{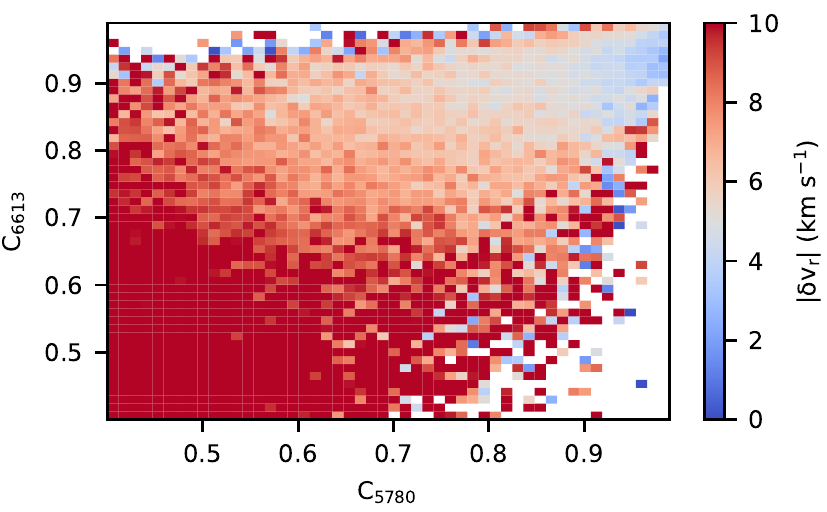}
\caption[f1]{Verification of correction of radial velocities of DIB clouds in different lines of sight. The highest values of correlation coefficients ($C$) for the DIBs at 5780 and 6613 \AA\ correspond to small differences in their inferred radial velocities ($|\delta v_r|$).}
\label{fig:Figure 5a}
\end{center}
\end{figure}

With the assumption that all DIBs in one LoS have the same radial velocity, we can correct ISM spectra of other spectral arms, using the values of the radial velocity shift obtained from DIB 5780. To confirm that this assumption is reasonable, we can use DIB 6613 in the red spectral arm which has similar strength to DIB 5780 and compare its measured radial velocities to the ones of DIB 5780. If the obtained radial velocity shifts are comparable, we can safely generalize the shift to other spectral arms. In Fig. \ref{fig:Figure 5a} we show that the average absolute difference of radial velocities between DIB 5780 and DIB 6613 are close to 0 km/s for spectra whose highest correlation coefficient (C) is greater than 0.8. We use C = 0.8 as the lower limit when combining ISM spectra into an averaged ISM spectrum. With this limit we perform a radial velocity shift for 71,626 ISM spectra at every spectral arm and then combine them based on the photometric values of reddening $E(\mathrm{G_{BP}-G_{RP}})$ provided by the Gaia DR3 archive \citep{GaiaDR3-2022, Gaia2016Prusti}. The $E(\mathrm{G_{BP}-G_{RP}})$ is converted to $E(\mathrm{B-V})$ using the calculated ratios between the extinction in each band and the reddening $E(\mathrm{B-V})$ as a function of the effective temperature as described in \citet{Montalto2021}. The estimates of $E(\mathrm{G_{BP}-G_{RP}})$ are used to map the total Galactic extinction \citep{Andrae2022} and are in good agreement with Planck data for $A_0 <$ 4 mag \citep{Delchambre2022} and with the extinction from \citet{SFD2011}. \citet{Schultheis2022} report good agreement with $E(\mathrm{G_{BP}-G_{RP}})$ and the $EW$ of DIB 8620 measured from Gaia Radial Velocity Spectrometer \citep{Blanco2022} spectra.

We need reddening values only for the purpose of dividing ISM spectra into six reddening bins between 0.1 and 0.7 mag containing 5981, 7847, 6616, 5440, 3928 and 3303 ISM spectra, respectively. Spectra in each bin are then combined to the averaged (mean) ISM spectrum. The mean values of reddening $E(\mathrm{B-V})$ of the averaged ISM spectra are: 0.15, 0.25, 0.35, 0.45, 0.55 and 0.65 mag.

Other maps of dust reddening, i.e.\ SFD \citep{SFD2011}, Planck 2013 \citep{Planck2014} and Stilism \citep{Stilism2014, Capitanio2017, Lallement2018} have also been considered. SFD and Planck 2013 provide only asymptotic values of reddening. Stilism on the other hand is a tridimensional map that provides reliable reddening estimates for targets at a distance of up to 2 kpc. Unfortunately, $\sim$23$\%$ of our targets (with C > 0.8) lie at a distance greater than 2 kpc, thus values of reddening at 2 kpc would have to be extrapolated for such targets. We test how Stilism compares to Gaia reddening and find that for a sample of stars at a distance of less than 2 kpc ($\sim$46,000 targets) the difference in reddening is $\sim$0.06 mag. While the difference is low our preferred reddening comes from Gaia as it is available for all targets including those beyond 2 kpc.

\subsubsection{Finding DIBs in the averaged ISM spectra} \label{Finding DIBs}

We find the absorption lines in the averaged ISM spectra (six spectra for each spectral arm) that show a positive correlation between the depth of the absorption line and the amount of reddening. We call them the DIB candidates.

The averaged ISM spectra contain a large number of absorption lines (see top panel in Fig. \ref{fig:Figure 6}, \ref{fig:Figure 7} and \ref{fig:Figure 8}). Note that the infra-red spectral band was discarded, as the quality of the averaged ISM spectra is poor due to a large number of strong air glow emission lines present in the spectra \citep{GALAH_reduc_pipeline_2017}. The averaged ISM spectra can also contain artifacts originating from incorrect reduction of spectra, coming mainly from peaks of H$\alpha$ and H$\beta$ absorption, telluric lines and interstellar emission lines. The latter can be present in some spectra and may have little correlation with interstellar reddening or with DIB clouds. In this regard we have to be careful when selecting an absorption feature in the averaged ISM spectrum as a DIB candidate. 

\begin{figure}
\begin{center}
\includegraphics[scale=1]{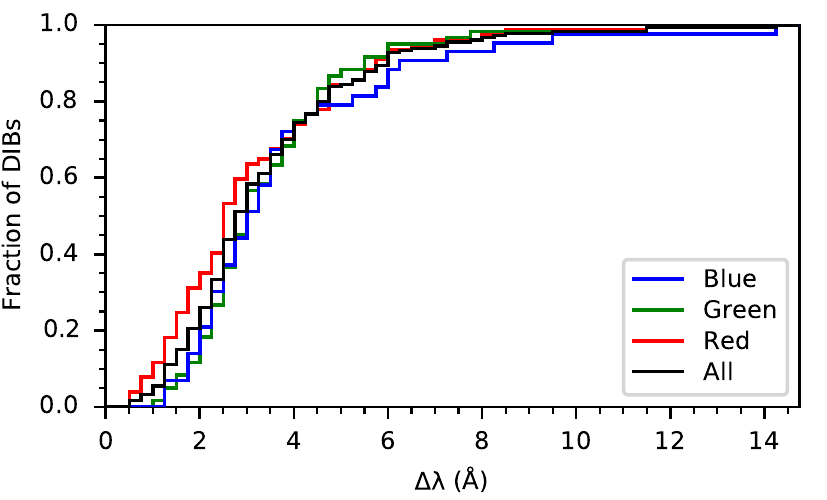}
\caption[f1]{The cumulative histogram of distances ($\Delta \lambda$) between the neighbouring DIBs for blue, green, red and all bands. The distances are computed using central wavelengths of DIBs listed in the catalogue, see Appendix \ref{sec:appendix_B}.}
\label{fig:Figure 5d}
\end{center}
\end{figure}

We find the DIB candidates using a pixel-wise correlation of the flux of the averaged ISM spectra with the amount of reddening. The criterion for detection is that the Pearson correlation coefficient is greater than 0.5. This alone is not sufficient for identification. At each pixel we calculate linear least-squares regression using the \texttt{scipy.stats.linregress} package \citep{ScipyStats} for relation of the $E\mathrm{(B-V)}$ against the flux. We obtain the slope and the intercept of the regression line for each pixel in the spectrum, see the third and the fourth panel from the top in figures in Appendix \ref{sec:appendix_A}. We use \texttt{scipy.signal.find$\_$peaks} \citep{ScipyStats} package to find peaks in the correlation coefficient, slope and offset plot, where we set the distance and the width between the neighbouring peaks, depending on the sampling size of the spectral arm. The minimal distance between two DIB candidates is set to $\sim$0.75 \AA~, which corresponds to a distance of 13 to 19 pixels, depending on the spectral arm. The choice of the minimal distance proves to be suitable, as a vast majority (80$\%$) of our DIBs in all bands are more than 2 \AA~apart and 97$\%$ of DIBs are more than 1 \AA~apart, see Fig. \ref{fig:Figure 5d}. Note that three DIBs in the red band fall in the bin below 0.75 \AA, with the lowest one at the distance of 0.65 \AA. This is possible because the central wavelengths of the fitted DIBs shift slightly from their initial positions. The dispersion of the DIB candidate is set to 3 pixels which corresponds to the lower limit of the width of the DIB, $\sim$0.15 \AA~wide ($FWHM \sim$0.35 \AA). The \texttt{scipy.signal.find$\_$peaks} \citep{ScipyStats} package finds all local maxima by simple comparison of the neighbouring values.

We then check if the peaks in the slope plot match with either the correlation coefficient plot or the offset plot. Their average positions are used as an initial guess of central wavelengths of the DIB candidates. The peak in the slope plot alone does not suffice the condition of the DIB candidate. The positions of the DIB candidates are then used in the fitting of the averaged ISM spectrum (see Section \ref{Measuring parameters}). We identify 395 DIB candidates, 123 in blue, 123 in green and 149 in red band. Additionally, we divide DIB candidates into two groups, the first having DIBs whose depth increases monotonically with increasing reddening and the second, where the latter is not the case. The depth of the DIB candidate increases with increasing reddening but at some reddening it could have a depth lower than its predecessor. The reason for this could be a weaker dependence of a specific DIB to the amount of interstellar reddening.    

\subsubsection{Measuring the DIB's parameters} \label{Measuring parameters}

The final step of our procedure is to fit the DIBs in six averaged ISM spectra with a parametric model containing the following parameters: central wavelength, half-width, asymmetry and equivalent width. The averaged ISM spectra have mean reddenings of 0.15, 0.25, 0.35, 0.45, 0.55 and 0.65 mag. We assume that the $EW$ of a DIB changes linearly with $E(\mathrm{B-V})$, as described by the following function:
\begin{equation}
     EW(E(\mathrm{B-V}); k, n) = kE(\mathrm{B-V})+n,
\label{eq:eq3}
\end{equation}
where $k = \Delta EW / \Delta E(\mathrm{B-V})$ is the slope and $n = EW(E$ $\mathrm{(B-V) = 0})$ is the intercept. The choice of linear relation between the equivalent width and the reddening is discussed in Section \ref{Results}.

We use central wavelengths of previously obtained DIB candidates as initial positions for the fitting procedure. The following is done by moving from one DIB candidate to another in the order of increasing wavelength, while fitting a sum of five DIB candidates (the one in question and its two closest DIB candidates to the left and to the right), each with a Gaussian function of the form:
\begin{equation}
    \Phi(\lambda; EW, \lambda_c, \sigma) = \frac{EW}{\sqrt{2\pi}\sigma}\exp\bigg({\frac{-(\lambda- \lambda_c)^2}{2\sigma^2}}\bigg),
\label{eq:eq1}
\end{equation}
where $\sigma$ is the half-width, $\mathrm{\lambda_c}$ is the central wavelength and $EW$ is the equivalent width of the DIB, and must be a linear function of reddening. The reason of fitting over five lines together is to improve the continuum fit which is not flat in many cases. The neighbouring DIB candidates can greatly complicate the fit of the DIB candidate in question, due to line overlap. Thus, the neighbouring DIB candidates can in a sense be thought of as the continuum around the DIB candidate in question. We test the quality of the fit across the whole spectrum in the case of the sum of five DIB candidates by computing chi-square values for each reddening bin. The chi-square values are of the order of $10^{-3}$. We repeat the computation in the case of the sum of nine DIB candidates and obtain similar results. The differences between these methods are $\sim$10$^{-4}$. Adding more DIB candidates does not change the quality of the continuum fit. Additionally, we introduce the asymmetry ($\mathrm{\gamma}$) of the profile in the width as a function of wavelength:
\begin{equation}
     \sigma(\lambda; \lambda_c, \gamma) = \frac{2\sigma}{1 + \exp{(\gamma(\lambda-\lambda_c))}},
\end{equation}
in order to fit DIBs whose shapes are best described with an asymmetric Gaussian function. The asymmetry is a free parameter only for strong and some weak but skewed DIBs that are selected after a visual inspection of the averaged ISM spectrum. Otherwise the asymmetry remains fixed to zero. 

We fit the sum of five Gaussians to the averaged ISM spectrum using the least squares minimization method \texttt{minimize()} in the \texttt{lmfit} package \citep{lmfit} to obtain the best fit parameters and their uncertainties estimated from the covariance matrix. The fit is performed simultaneously for all averaged ISM spectra in the band, meaning that we minimize the combined residual of the fit for six averaged ISM spectra. The result of fitting is a set of 5 fitting parameters ($\lambda_c$, $\sigma$, $\gamma$, $k$, $n$) per fitted DIB.   

In some cases DIBs cannot be fitted effectively with the described model, because they are either too narrow or their shape resembles a blend of a more prominent narrow DIB stacked on top of a broader and shallower DIB. Therefore we manually calculate the $EW$ and further obtain the slope ($k$) and the intercept ($n$) in the $EW$ vs $E(\mathrm{B-V})$ relation, see Eq. \ref{eq:eq3}. We measure the half-width and the central wavelength by hand and keep them fixed, while the asymmetry is set to zero. The parameter uncertainties for $k$ and $n$ are estimated through ordinary least squares method, while the remaining parameter uncertainties are estimated using the covariance matrix in the uniform noise case for the one-dimensional Gaussian profile \citep{Hagen07}. 

Also note that we do not use all DIB candidates in the fitting procedure, because we want to use the least number of DIBs possible to still obtain a well fitted averaged ISM spectrum. This is done by initially fitting all DIB candidates. Then, while leaving
DIBs from the first group of DIB candidates intact, see Sec. \ref{Finding DIBs}, we remove DIBs from the second group, one by one until the fitting residual reaches minimum.     

We create tables of DIBs (see Appendix \ref{sec:appendix_B}) and arrange them into three quality classes based on their relative parameter uncertainties (see Table \ref{tab:table_criterion}): certain (\texttt{+}), probable (\texttt{$\circ$}) and possible (\texttt{-}). The choice of quality is set arbitrarily, depending on the relative uncertainties of the equivalent widths ($EW$) and the slopes ($k$) of the strong DIBs in each band. Additional criterion is a lower limit on the DIB's amplitude ($A$), which equals 3$\sigma$, where $\sigma$ is a standard deviation of the spectral flux measured for each spectral band. All three criteria differ slightly between the bands, as shown in Table \ref{tab:table_criterion}.

\begin{table}
\centering
\caption{The choice of the DIB quality (Q) is based on its amplitude ($A$), and relative uncertainties of the slope ($k$) and the equivalent width ($EW$). The DIB quality is split in three classes: certain (\texttt{+}), probable (\texttt{$\circ$}) and possible (\texttt{-}).}
\label{tab:table_criterion}
\begin{tabular}{l c c c c c l}
\hline
Band & Q & $A (\%)$ & & $\sigma (k)/k$ & & $\sigma(\mathrm{ EW})/{\mathrm EW}$\\\hline\hline
Blue & \texttt{+} & > 0.35 & $\&$ & < 0.5 & $\&$ & < 0.3\\
& \texttt{$\circ$} & > 0.35 & $\&$ & < 0.5 & $\&$ & [0.3, 0.5)\\
& \texttt{-} & > 0 & $\&$ & ($\geq$ 0.5 & | & $\geq$ 0.5)\\
\hline
Green & \texttt{+} & > 0.25 & $\&$ & < 0.3 & $\&$ & < 0.3\\
& \texttt{$\circ$} & > 0.25 & $\&$ & < 0.3 & $\&$ & [0.3, 0.5)\\
& \texttt{-} & > 0 & $\&$ & ($\geq$ 0.3 & | & $\geq$ 0.5)\\
\hline
Red & \texttt{+} & > 0.20 & $\&$ & < 0.3 & $\&$ & < 0.3\\
& \texttt{$\circ$} & > 0.20 & $\&$ & < 0.3 & $\&$ & [0.3, 0.5)\\
& \texttt{-} & > 0 & $\&$ & ($\geq$ 0.3 & | & $\geq$ 0.5)\\
\hline
\end{tabular}
\end{table}

We compare the measured DIBs with the ones reported in the literature: \citetalias{HobbsYork2009}, \citetalias{Tuairisg2000}, \citetalias{Jenniskens1994}, \citetalias{Fan2019}, \citetalias{Sonnentrucker2018}, \citetalias{GalazutdinovMusaev2000} and \citetalias{Weselak2000}. We declare a match if the reported central wavelength from any aforementioned catalogue is within 1 \AA~from the measured central wavelength. However, some exceptions are allowed in the case of very broad DIBs, as the position of the central wavelength can vary substantially (more than 1 \AA). The first five of the mentioned catalogues also allow for comparison between the DIB's $FWHM$. A short sample of our DIB catalogue is shown in Table \ref{tab:table_dibs_example}.

\begin{table*}
\centering
\caption{A short sample of our DIB catalogue. The quality (Q) of each DIB is divided in three classes based on the relative uncertainty of its parameters: certain (\texttt{+}), probable (\texttt{$\circ$}) and possible (\texttt{-}). For each DIB, one can find the value of the central wavelength ($\mathrm{\lambda_C}$), the width ($FWHM$), the asymmetry ($\gamma$) of the Gaussian fit of the DIB where available, the slope and the intercept of the regression line in the relation of the equivalent width against the reddening. $\lambda_{\mathrm{REF}}$ ($FWHM_{\mathrm{REF}}$) shows cross-reference values for DIBs based on their central wavelength and if available their $FWHM$. The comment denoted with ($\dagger$) marks the DIBs that were fitted manually, and ($\ddag$) marks new DIBs that have no correspondence in the reference DIB catalogues.}
\label{tab:table_dibs_example}
\begin{tabular}{l c c c c c c l}
\hline
Q & $\lambda_{C}$ & $\lambda_{\mathrm{REF}}$ ($FWHM_{\mathrm{REF}}$) & $FWHM$ & $\gamma$ & $\Delta EW$ / $\Delta E(\mathrm{B-V})$ & $EW(E\mathrm{(B-V) = 0})$ & Comment\\
& (\AA) & (\AA) & (\AA) & & ($\mathrm{m}$\AA\ / $\mathrm{mag}$) & ($\mathrm{m}$\AA) &\\ \hline\hline
\texttt{+} & 6597.37 $\pm$ 0.09 & 6597.43 (0.69)$^{\mathrm{a}}$, 6597.31$^{\mathrm{b}}$, ... & 0.77 $\pm$ 0.18 & & 3.20 $\pm$ 0.22 & 5.70 $\pm$ 0.39 & \\
\texttt{$\circ$} & 6598.57 $\pm$ 0.13 & & 0.79 $\pm$ 0.13 & & 5.83 $\pm$ 0.62 & 0.07 $\pm$ 0.29 & $\dagger$, $\ddag$\\
\texttt{$\circ$} & 6600.28 $\pm$ 0.60 & 6600.09 (0.65)$^{\mathrm{a}}$, 6599.99 (0.50)$^{\mathrm{f}}$ & 1.80 $\pm$ 2.24 & -3.00 $\pm$ 9.75 & 9.60 $\pm$ 2.59 & -0.82 $\pm$ 1.22 & \\
\texttt{+} & 6604.11 $\pm$ 0.89 & & 3.31 $\pm$ 1.71 & & 16.16 $\pm$ 2.00 & 0.20 $\pm$ 0.94 & $\ddag$\\
\texttt{$\circ$} & 6606.85 $\pm$ 0.23 & 6607.09 (0.60)$^{\mathrm{a}}$, 6607.10 (0.53)$^{\mathrm{f}}$ & 1.27 $\pm$ 0.23 & & 5.61 $\pm$ 1.83 & 1.10 $\pm$ 0.86 & $\dagger$\\
\texttt{$\circ$} & 6611.32 $\pm$ 0.32 & 6611.06 (1.06)$^{\mathrm{a}}$, 6611.14 (1.39)$^{\mathrm{f}}$ & 1.41 $\pm$ 0.62 & & 6.90 $\pm$ 1.81 & 0.72 $\pm$ 0.85 & \\
\texttt{+} & 6613.66 $\pm$ 0.01 & 6613.70 (1.08)$^{\mathrm{a}}$, 6613.56$^{\mathrm{b}}$, ... & 1.02 $\pm$ 0.02 & -0.57 $\pm$ 0.08 & 130.37 $\pm$ 4.38 & 31.90 $\pm$ 2.06 & \\
\texttt{-} & 6616.34 $\pm$ 0.67 & 6616.12 (0.61)$^{\mathrm{f}}$ & 3.71 $\pm$ 1.20 & -0.03 $\pm$ 0.42 & 1.64 $\pm$ 4.72 & 6.47 $\pm$ 2.23 & \\
\hline
\end{tabular}
\begin{tablenotes}
      \small
      \item Cross-reference with a DIB noted in \texttt{a}: \citetalias{HobbsYork2009}; \texttt{b}: \citetalias{GalazutdinovMusaev2000}, \texttt{f}: \citetalias{Fan2019}, ...
      
    \end{tablenotes}
\end{table*}

\section{Results and discussion} \label{Results}
\begin{landscape}
\begin{figure}
\includegraphics[scale=0.94]{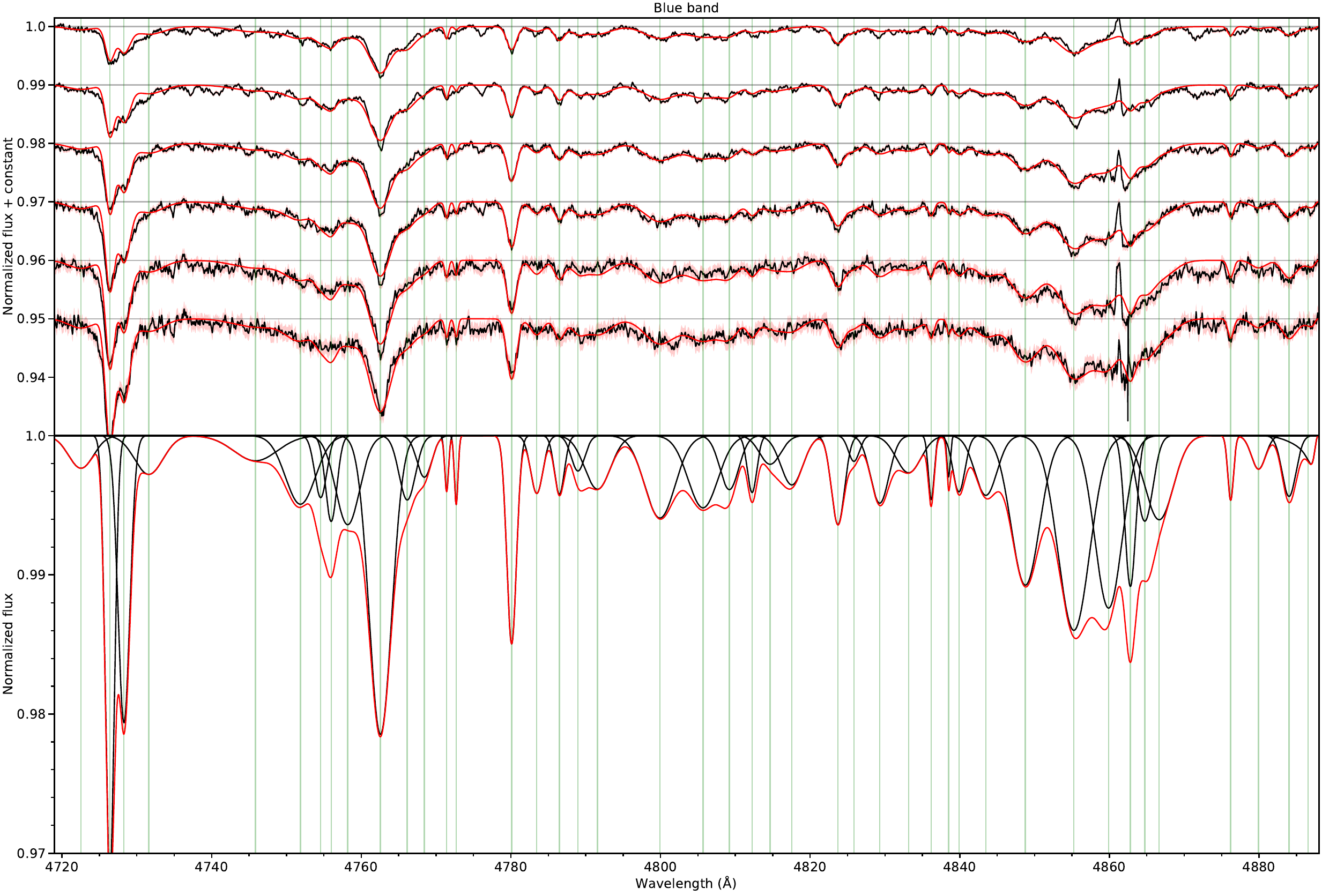}
\caption{DIBs in the blue band. The top panel shows a sequence of averaged ISM spectra (black) with mean reddenings   increasing from $E\mathrm{(B-V)} = 0.15$~mag (top) to 0.65~mag (bottom) in steps of 0.1~mag. The sum of the Gaussian functions with measured parameters (red) is scaled to the reddening of the averaged ISM spectrum and plotted over it. Green vertical lines show the measured central wavelengths of DIBs. The bottom panel shows individual DIBs (black) at $E\mathrm{(B-V)}$ = 1 mag, and the red line shows their sum.}
\label{fig:Figure 11}
\end{figure}
\end{landscape}

\begin{landscape}
\begin{figure}
\includegraphics[scale=0.94]{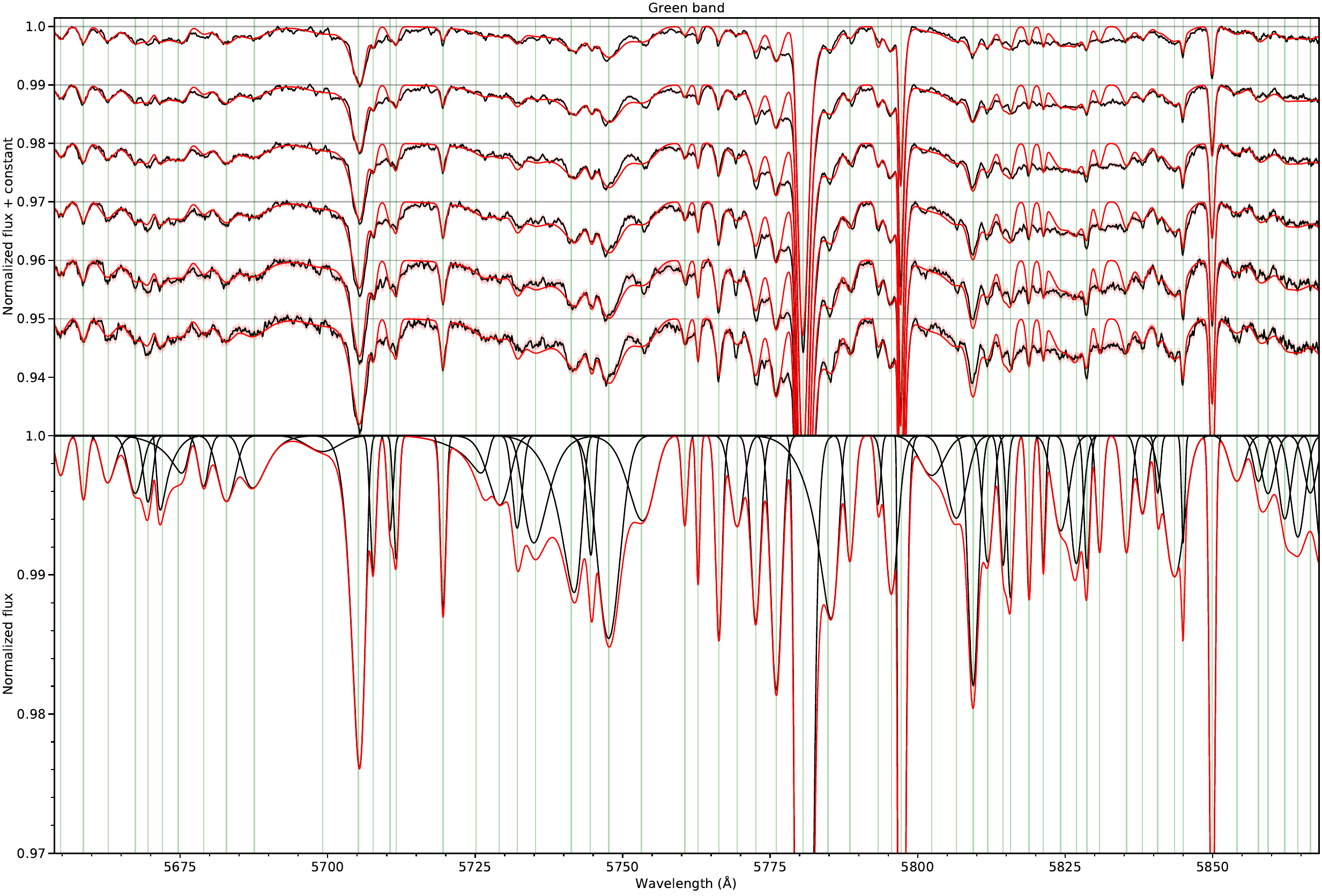}
\caption{As Fig. \ref{fig:Figure 11} but for the green band.}
\label{fig:Figure 12}
\end{figure}
\end{landscape}

\begin{landscape}
\begin{figure}
\includegraphics[scale=0.94]{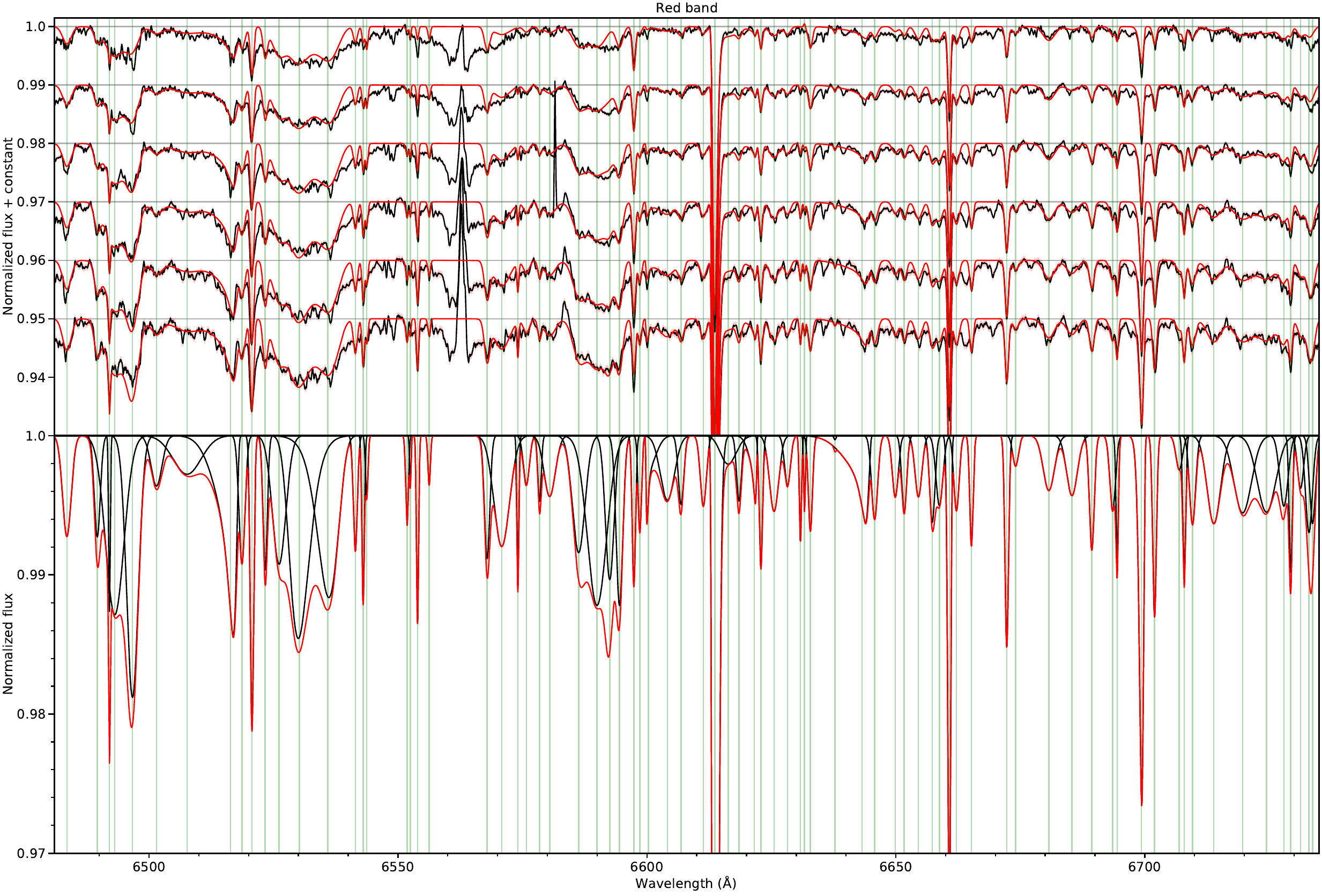}
\caption{As Fig. \ref{fig:Figure 11} but for the red band.}
\label{fig:Figure 13}
\end{figure}
\end{landscape}

Averaged ISM spectra with reddening of $E(\mathrm{B-V})$ = 0.15, 0.25, 0.35, 0.45, 0.55 and 0.65 mag  are presented in Figs. \ref{fig:Figure 11}, \ref{fig:Figure 12}, and \ref{fig:Figure 13}. The bottom panels of these figures show the synthetic DIB spectrum scaled to the reddening of $E\mathrm{(B-V)} = 1$ mag that was built using the tabulated DIB parameters, see Appendix \ref{sec:appendix_B}. Note that wavelengths around the H$\alpha$ and H$\beta$ lines show substantial residuals due to weak emission features which may vary from star to star and are not reproduced by our stellar spectral templates.

We detect a total of 183 DIBs (44 in blue band, 61 in green band, 77 DIBs and 1 interstellar absorption line of Li I in red band). The catalogue (see Appendix \ref{sec:appendix_B}) contains central wavelengths and widths of the Gaussian fits of the DIBs, asymmetry indices in the profile of the width as a function of wavelength, as well as slopes and intercepts of the regression line in the relation of DIBs' equivalent widths against the reddening. We discover 31 DIBs (17$\%$ of all DIBs detected) that are not reported in DIB catalogues in the literature, see Table \ref{tab:table_new_dibs}. DIBs are divided into three quality classes, based on their relative uncertainty of parameters, as explained in Section \ref{Measuring parameters}. Among the new discoveries 17 DIBs are categorized as certain (\texttt{+}) and 14 as probable (\texttt{$\circ$}). 33 possible (\texttt{-}) DIBs that are not reported in DIB catalogues are excluded from the category of newly discovered DIBs. We identify 119 DIBs that are already in the literature, 73 are categorized as certain, 28 as probable and 18 as possible. 26 DIBs are fitted using asymmetry ($\gamma$) as a free parameter. We cannot find and confirm 51 DIBs noted in the literature (see Table \ref{tab:table_dibs_notreported}), which should lie in our observed bands.

\begin{table}
\begin{center}
\centering
\caption{Table of 31 new DIBs detected in our survey that are not reported in DIB catalogues in the literature. The quality (Q) of each DIB is divided in two classes based on the relative uncertainty of its parameters: certain (\texttt{+}) and probable (\texttt{$\circ$}). For each DIB, one can find the value of the central wavelength ($\mathrm{\lambda_C}$), the width ($FWHM$), the slope and the intercept of the regression line in the relation of equivalent width against the reddening.}
\label{tab:table_new_dibs}
\begin{tabular}{c c c c c}
\hline
Q & $\lambda_{C}$ & $FWHM$ & $\Delta EW$ / & $EW$\\
& & & $\Delta E(\mathrm{B-V})$ & $(E\mathrm{(B-V) = 0})$ \\
& (\AA) & (\AA) & ($\mathrm{m}$\AA\ / $\mathrm{mag}$) & ($\mathrm{m}$\AA)
\\\hline\hline
\texttt{$\circ$} & 4751.87 $\pm$ 0.31 & 4.68 $\pm$ 0.59 & 17.66 $\pm$ 1.46 & 6.91 $\pm$ 0.63 \\
\texttt{$\circ$} & 4791.58 $\pm$ 0.40 & 4.04 $\pm$ 0.79 & 12.91 $\pm$ 4.81 & 3.66 $\pm$ 2.08 \\
\texttt{+} & 4799.92 $\pm$ 0.25 & 4.76 $\pm$ 0.56 & 23.50 $\pm$ 5.17 & 6.54 $\pm$ 2.24 \\
\texttt{$\circ$} & 4805.68 $\pm$ 0.77 & 4.68 $\pm$ 0.62 & 19.94 $\pm$ 4.88 & 5.93 $\pm$ 2.12 \\
\texttt{$\circ$} & 4829.30 $\pm$ 0.41 & 2.89 $\pm$ 0.81 & 14.86 $\pm$ 5.40 & 0.09 $\pm$ 2.34 \\
\texttt{$\circ$} & 4843.49 $\pm$ 0.38 & 3.49 $\pm$ 0.74 & 13.95 $\pm$ 2.20 & 2.02 $\pm$ 0.95 \\
\texttt{+} & 4848.77 $\pm$ 0.14 & 4.46 $\pm$ 0.33 & 46.05 $\pm$ 3.41 & 5.00 $\pm$ 1.48 \\
\texttt{+} & 4855.25 $\pm$ 0.12 & 4.95 $\pm$ 0.24 & 59.35 $\pm$ 4.67 & 14.36 $\pm$ 2.02 \\
\texttt{+} & 4859.89 $\pm$ 0.01 & 4.28 $\pm$ 0.38 & 56.75 $\pm$ 7.11 & -0.35 $\pm$ 3.08 \\
\texttt{+} & 4862.81 $\pm$ 0.14 & 1.72 $\pm$ 0.26 & 19.07 $\pm$ 1.44 & 0.79 $\pm$ 0.63 \\
\texttt{$\circ$} & 4864.73 $\pm$ 0.24 & 2.54 $\pm$ 0.48 & 14.32 $\pm$ 2.10 & 2.28 $\pm$ 0.91 \\
\texttt{+} & 4866.65 $\pm$ 0.07 & 4.47 $\pm$ 0.11 & 31.30 $\pm$ 7.82 & -2.50 $\pm$ 3.39 \\
\texttt{$\circ$} & 5658.61 $\pm$ 0.44 & 1.47 $\pm$ 0.85 & 3.29 $\pm$ 0.36 & 3.96 $\pm$ 0.15 \\
\texttt{$\circ$} & 5662.80 $\pm$ 2.02 & 2.82 $\pm$ 1.43 & 3.30 $\pm$ 0.60 & 6.86 $\pm$ 0.26 \\
\texttt{$\circ$} & 5667.39 $\pm$ 0.63 & 2.81 $\pm$ 1.20 & 4.92 $\pm$ 0.72 & 7.50 $\pm$ 0.31 \\
\texttt{$\circ$} & 5687.58 $\pm$ 1.10 & 4.80 $\pm$ 2.04 & 10.30 $\pm$ 2.57 & 8.76 $\pm$ 1.11 \\
\texttt{+} & 5729.04 $\pm$ 1.09 & 4.41 $\pm$ 1.95 & 17.79 $\pm$ 1.30 & 5.56 $\pm$ 0.56 \\
\texttt{+} & 5824.25 $\pm$ 0.53 & 3.23 $\pm$ 1.03 & 14.52 $\pm$ 1.34 & 9.03 $\pm$ 0.58 \\
\texttt{+} & 5826.89 $\pm$ 0.46 & 2.46 $\pm$ 0.88 & 19.70 $\pm$ 0.49 & 4.34 $\pm$ 0.21 \\
\texttt{+} & 5864.43 $\pm$ 0.77 & 2.89 $\pm$ 1.50 & 20.86 $\pm$ 2.27 & 1.54 $\pm$ 0.98 \\
\texttt{+} & 5868.36 $\pm$ 0.78 & 2.16 $\pm$ 1.49 & 18.58 $\pm$ 2.66 & -0.59 $\pm$ 1.15 \\
\texttt{+} & 6507.63 $\pm$ 1.45 & 6.59 $\pm$ 2.78 & 13.99 $\pm$ 2.72 & 5.52 $\pm$ 1.28 \\
\texttt{+} & 6526.11 $\pm$ 0.01 & 3.17 $\pm$ 0.47 & 17.49 $\pm$ 1.81 & 13.69 $\pm$ 0.86 \\
\texttt{+} & 6541.43 $\pm$ 0.12 & 0.94 $\pm$ 0.12 & 6.67 $\pm$ 0.50 & 1.63 $\pm$ 0.23 \\
\texttt{$\circ$} & 6551.82 $\pm$ 0.12 & 0.62 $\pm$ 0.12 & 4.44 $\pm$ 0.85 & -0.18 $\pm$ 0.40 \\
\texttt{+} & 6574.08 $\pm$ 0.07 & 0.46 $\pm$ 0.07 & 5.91 $\pm$ 0.77 & -0.86 $\pm$ 0.36 \\
\texttt{$\circ$} & 6575.78 $\pm$ 0.66 & 1.40 $\pm$ 1.28 & 4.62 $\pm$ 1.31 & 0.66 $\pm$ 0.62 \\
\texttt{$\circ$} & 6578.43 $\pm$ 0.46 & 0.79 $\pm$ 0.88 & 4.01 $\pm$ 0.69 & -0.01 $\pm$ 0.32 \\
\texttt{+} & 6586.28 $\pm$ 0.39 & 2.94 $\pm$ 0.67 & 20.25 $\pm$ 5.91 & 6.04 $\pm$ 2.79 \\
\texttt{$\circ$} & 6598.57 $\pm$ 0.13 & 0.79 $\pm$ 0.13 & 5.83 $\pm$ 0.62 & 0.07 $\pm$ 0.29 \\
\texttt{+} & 6604.11 $\pm$ 0.89 & 3.31 $\pm$ 1.71 & 16.16 $\pm$ 2.00 & 0.20 $\pm$ 0.94 \\
\hline
\end{tabular}
\end{center}
\end{table}

We present some statistical properties of DIBs in Table \ref{tab:table_stats}. We obtain minimum, maximum and median values of $FWHM$ and $EW$ at $E(\mathrm{B-V}) = 1$ mag of all DIBs in the catalogue. We calculate the median amplitude (depth) of DIBs described by the Gaussian function, using the following equation:
\begin{equation}
    A = \frac{2.355EW}{FWHM\sqrt{2\pi}},
    \label{eq:eq2}
\end{equation}
where $FWHM$ and $EW$ are median values. We find that DIBs in our survey have median $FWHM$ of $\sim$1.87 \AA~and median $EW$ of $\sim$11.86 m\AA, which gives an estimate of median amplitude ($A$) of $\sim$0.59$\%$ from the continuum at 1. We note that DIBs in the red band are the narrowest ($\sim$1.38 \AA) and DIBs in the blue band are the broadest ($\sim$2.44 \AA). The strongest ($EW$) DIBs are seen in the green band ($\sim$14.56 m\AA) and the weakest in the red band ($\sim$9.05 m\AA). Consequently, the most prominent (in amplitude) DIBs can be seen in the green band ($\sim$0.64$\%$) and the least prominent DIBs in the blue band ($\sim$0.48$\%$). Our narrowest DIB is at 6552.47 \AA\ with $FWHM$ = 0.35 \AA, and the broadest at 4745.87 \AA\ with $FWHM$ = 6.83 \AA. The strongest DIB is at 5780.59 \AA\ with $EW$ = 357.49 m\AA\ and the weakest is at 6552.47 \AA\ with $EW$ = 1.30 m\AA, in both cases normalized to a colour excess of $E(\mathrm{B-V}) =$ 1 mag.  

\begin{table}
\begin{center}
\centering
\caption{Table of central wavelengths of DIBs from the literature (REF) that are not fitted to the averaged ISM spectra but are listed among some of the DIB candidates (CDT) obtained by our DIBs selection method. Note that 11 DIBs from the literature were overlooked by our DIB seeking method.}
\label{tab:table_dibs_notreported}
\begin{tabular}{l l l l}
\hline
$\lambda_{\mathrm{REF}}$ (\AA) & $\lambda_{\mathrm{CDT}}$ (\AA) & $\lambda_{\mathrm{REF}}$ (\AA) & $\lambda_{\mathrm{CDT}}$ (\AA)
\\\hline\hline
4734.77$^{\mathrm{f}}$ & 4734.92 & 5855.63$^{\mathrm{b}}$ & 5855.69\\
4761/4764$^{\mathrm{f,g}}$ & & 6498.01$^{\mathrm{a}}$ & 6497.75\\
4881$^{\mathrm{g}}$ & 4881.59 & 6513.76$^{\mathrm{a}}$ & 6513.53\\
4881.63$^{\mathrm{f}}$ & 4881.59 & 6513.78$^{\mathrm{f}}$ & 6513.53\\
5652.34$^{\mathrm{f}}$ & 5652.50 & 6534.54$^{\mathrm{a}}$ & 6534.77\\
5704$^{\mathrm{g}}$ & & 6548.99$^{\mathrm{f}}$ & 6549.18\\
5706.50$^{\mathrm{f}}$ & & 6549.09$^{\mathrm{a}}$ & 6549.18\\
5709.40$^{\mathrm{c}}$ & 5709.46 & 6572.93$^{\mathrm{f}}$ & 6572.82\\
5716.30$^{\mathrm{f}}$ & 5716.52 & 6573.04$^{\mathrm{a}}$ & 6572.82\\
5746.21$^{\mathrm{c}}$ & & 6624.93$^{\mathrm{a}}$ & 6625.57\\
5746.93$^{\mathrm{a}}$ & 5747.17 & 6635.66$^{\mathrm{a,f}}$ & 6635.66\\
5749.18$^{\mathrm{f}}$ & 5748.52 & 6639.32$^{\mathrm{f}}$ & 6639.44\\
5756.10$^{\mathrm{f}}$ & 5756.32 & 6639.42$^{\mathrm{a}}$ & 6639.44\\
5758.92$^{\mathrm{f}}$ & 5758.87 & 6644.33$^{\mathrm{c}}$ & 6644.66\\
5766.98$^{\mathrm{a}}$ & 5766.28 & 6657.47$^{\mathrm{a}}$ & 6657.32\\
5779$^{\mathrm{g}}$ & & 6664.05$^{\mathrm{a}}$ & \\
5779.58$^{\mathrm{f}}$ & & 6669.36$^{\mathrm{f}}$ & 6669.38\\
5842.23$^{\mathrm{b}}$ & & 6669.51$^{\mathrm{a}}$ & 6669.38\\
5842.38$^{\mathrm{c}}$ & & 6686.46$^{\mathrm{c}}$ & 6687.09\\
5842.49$^{\mathrm{d}}$ & & 6697.08$^{\mathrm{f}}$ & 6697.29\\
5843.30$^{\mathrm{c}}$ & 5843.64 & 6717.63$^{\mathrm{f}}$ & 6718.33\\
5843.42$^{\mathrm{d}}$ & 5843.64 & 6718.38$^{\mathrm{a}}$ & 6717.63\\
5843.69$^{\mathrm{c}}$ & 5843.64 & 6719.25$^{\mathrm{f}}$ & 6718.33\\
5845.67$^{\mathrm{d}}$ & 5844.99 & 6728.55$^{\mathrm{a,f}}$ & 6729.28\\
5848.37$^{\mathrm{d}}$ & & 6735.43$^{\mathrm{f}}$ & 6734.62\\
5855.62$^{\mathrm{f}}$ & 5855.69 & & \\
\hline
\end{tabular}
\begin{tablenotes}
      \small
      \item DIB noted in \texttt{a}: \citetalias{HobbsYork2009}; \texttt{b}: \citetalias{GalazutdinovMusaev2000}; \texttt{c}: \citetalias{Tuairisg2000}; \texttt{d}: \citetalias{Weselak2000}; \texttt{f}: \citetalias{Fan2019}; \texttt{g}: \citetalias{Sonnentrucker2018}.
    \end{tablenotes}
\end{center}
\end{table}

\begin{table}
\begin{center}
\centering
\caption{Table of minimum (Min), maximum (Max) and median (Med) values of the widths ($FWHM$) and strengths ($EW$) of the DIBs. The median amplitude ($A$) is calculated using Eq. \ref{eq:eq2}. Values are presented for DIBs in all bands and for DIBs in individual bands.}
\label{tab:table_stats}
\begin{tabular}{l c @{\hspace{0.2cm}} c @{\hspace{0.2cm}} c c @{\hspace{0.2cm}} c @{\hspace{0.2cm}} c c}
\hline
\multicolumn{1}{l}{Band} & \multicolumn{3}{c}{$FMHM$ (\AA)} & \multicolumn{3}{c}{$EW$ ($\mathrm{m}$\AA)} & \multicolumn{1}{c}{$A$ ($\%$)}\\
& Min & Max & Med & Min & Max & Med & Med\\\hline\hline
All & 0.35 & 6.83 & 1.87 & 1.30 & 357.49 & 11.86 & 0.59\\
Blue & 0.53 & 6.83 & 2.44 & 1.68 & 80.80 & 12.45 & 0.48\\
Green & 0.64 & 6.28 & 2.13 & 4.50 & 357.49 & 14.56 & 0.64\\
Red & 0.35 & 6.59 & 1.38 & 1.30 & 162.27 & 9.05 & 0.62\\
\hline
\end{tabular}
\end{center}
\end{table}

\begin{figure}
\begin{center}
\includegraphics[scale=1]{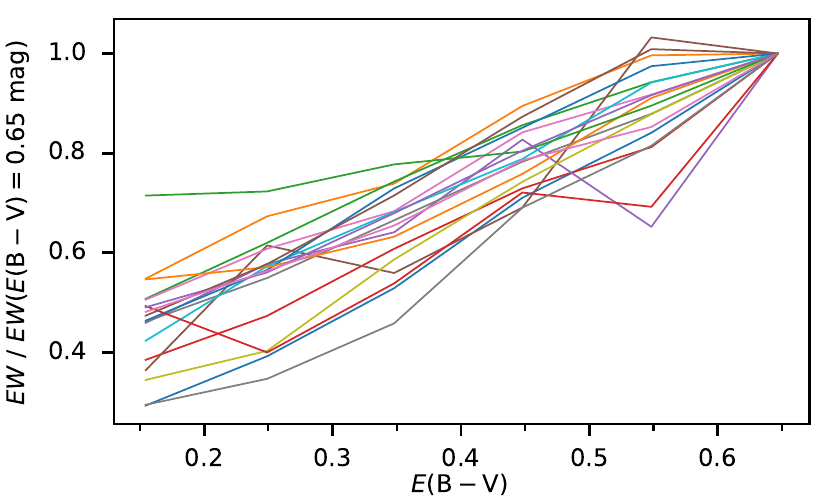}
\caption[f1]{The figure shows that the normalized equivalent width ($EW$) rises approximately linearly with the increasing amount of reddening $E(\mathrm{B-V})$. This is demonstrated with 18 (10$\%$ of all DIBs) different DIBs, six from each band. The central wavelengths of DIBs used in this example are the following: 4726.40, 4751.87, 4762.57, 4780.10, 4799.92, 4843.49, 5707.71, 5780.59, 5797.19, 5815.74, 5830.84, 5844.99, 6526.11, 6553.93, 6613.66, 6625.54, 6672.28, and 6713.88 \AA.}
\label{fig:Figure 5c}
\end{center}
\end{figure}

We find that the use of linear relationship between the equivalent width and the reddening is appropriate, see Fig. \ref{fig:Figure 5c}. The figure shows that the measured equivalent width rises approximately linearly with the increasing amount of reddening. Additionally, we calculate chi-square value for each DIB in our survey using two different functions, the first is our adopted linear function, the second is a quadratic function. The chi-square values for both functions have an average of about $10^{-3}$. The differences of chi-square values between the linear and the quadratic function are of the order of $10^{-4}$.

\begin{landscape}
\begin{figure}
\includegraphics[scale=0.94]{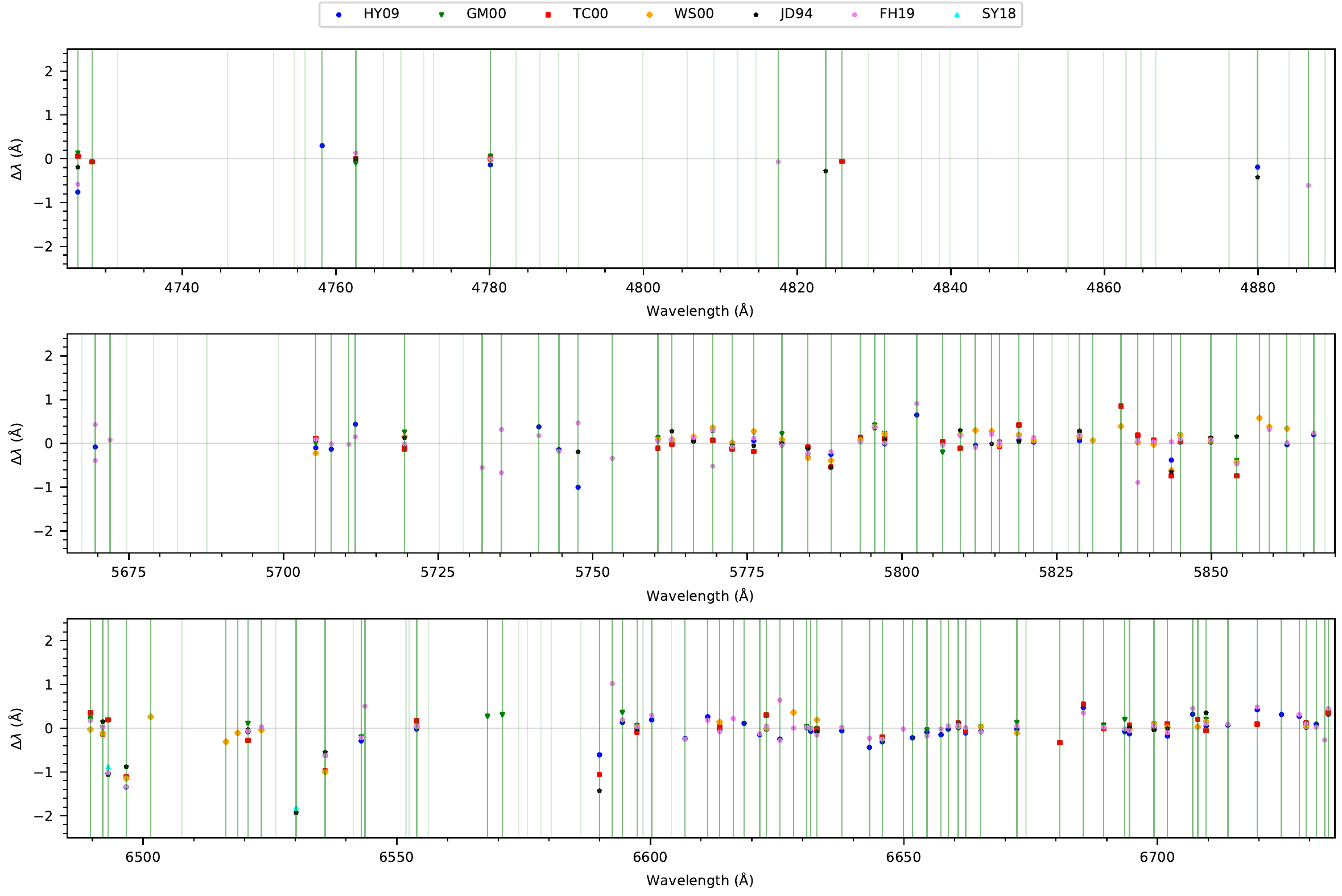}
\caption{Difference between wavelengths of the DIBs measured in this work and the values reported in the literature. Each vertical line (green) presents a measured central wavelength of a DIB from our catalogue. Lines in dark green coincide with central wavelengths of DIBs that have a cross-reference with DIBs reported in the literature. The horizontal line (gray) marks a perfectly matching case.}
\label{fig:Figure 10a}
\end{figure}
\end{landscape}

We compare DIBs from our catalogue with DIBs reported in the literature. In Fig. \ref{fig:Figure 10a} we show a difference between measured central wavelength and central wavelength reported in \citetalias{HobbsYork2009, GalazutdinovMusaev2000, Tuairisg2000, Weselak2000, Jenniskens1994, Fan2019, Sonnentrucker2018}. We find that a majority of DIBs lie close to zero difference, indicating that our measurements are consistent with other surveys. DIBs 4726.40, 5788.51, 5795.58, 5843.54, 5854.11, 6496.67, 6535.89 and 6685.38 all show greater offset when compared to other catalogues. The reason for inconsistency in these DIBs lies in their width, as all listed DIBs (except for DIB 4726.40 and 5788.51) have $FWHM$s greater than the median value of $FWHM$, as noted in Table \ref{tab:table_stats}. Broader DIBs tend to have less defined central wavelengths as their peaks are less prominent. We note that the DIB 5788.51 is located in the wings of an extremely strong DIB 5780.59, which may affect its position measurements in the literature.

\begin{landscape}
\begin{figure}
\includegraphics[scale=0.94]{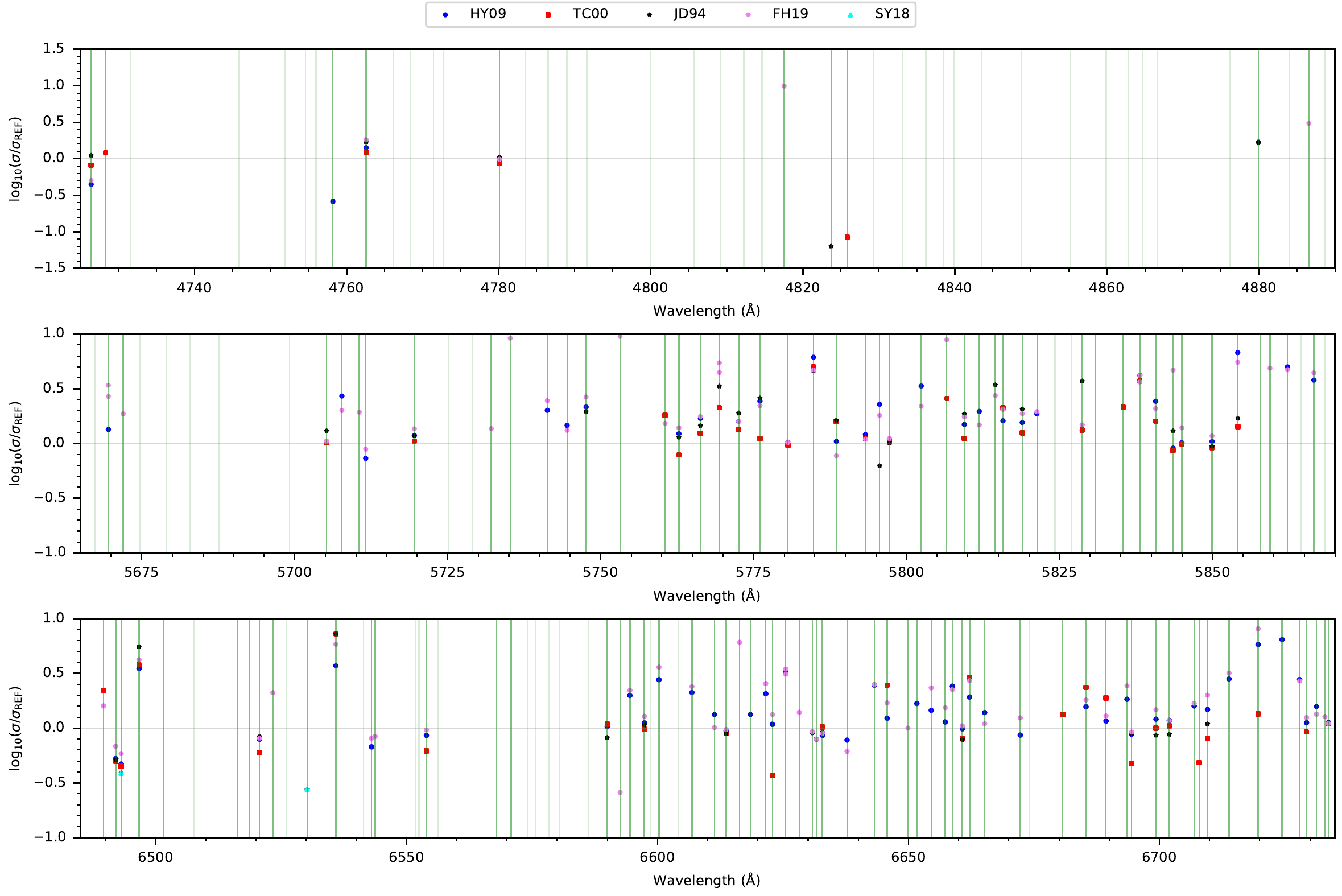}
\caption{Logarithmic ratio of DIB widths measured in this work and the corresponding values from the literature. Presentation scheme is the same as in Fig.\ \ref{fig:Figure 10a}.}
\label{fig:Figure 10b}
\end{figure}
\end{landscape}

Similarly, we compare ratios of our measured widths with the ones reported in \citetalias{HobbsYork2009, Tuairisg2000, Jenniskens1994, Fan2019, Sonnentrucker2018}, see Fig. \ref{fig:Figure 10b}. We present their ratios on a logarithmic scale. We notice a trend showing that our DIBs are generally broader than the ones in the literature, specifically in \citetalias{HobbsYork2009} and \citetalias{Fan2019}. This could be attributed to the fact that we average a few thousands of ISM spectra after shifting them for the velocity of a DIB cloud, which can contribute to broadening of DIBs even in spectra where correlation coefficient ($C$) of radial velocity shift is > 0.8, see Section \ref{RV corr}. All aforementioned surveys used a few high resolution spectra of hot stars with reddening $E(\mathrm{B-V}) > $ 0.1~mag, where this kind of broadening effect is negligible. The discrepancy is also caused by the different DIBs selection method, namely, in \citetalias{HobbsYork2009} the DIBs were found by inter-comparing three averaged high $SNR$ spectra by eye, while our method relies on the measurement of the correlation coefficient between the averaged ISM spectra with different amount of reddening, the slope and the intercept of the regression line in the relation of the $E(\mathrm{B-V})$ against the flux. Different selection methods can give different number of DIB candidates that build up the broad feature. Thus, one could fit the broad feature with fewer/more DIBs leading to estimated DIB widths that are broader/narrower.

An asymmetric Gaussian may not represent the true profile shape of broad DIBs and our limited resolving power does not permit to resolve the fine structure. A different profile that better fits the tails of the fine structure would therefore have to be implemented. The reported widths of broad DIBs may not be completely reliable, as these DIBs are ambiguous because they can be assembled from many narrower DIBs. This is demonstrated in Fig. \ref{fig:Figure 10b} by DIBs 5747.62, 5769.39, 5784.78, 5802.38, 5806.56, 5814.49, 5838.11, 5854.11, 5862.23, 5866.56, 6492.03, 6493.11, 6496.67, 6535.89, 6600.28, 6625.54, 6719.67 and 6724.45 which have our widths substantially different from values reported in the literature. All of these DIBs (except for DIB 5814.49, 5838.11 and DIB 6492.03) are broad. Additionally, we find that very broad DIBs ($\geq$ 6 \AA) listed in \citetalias{Sonnentrucker2018}, i.e. DIB 4761+4764, 5704 and 5779 are in our survey described by a multitude of narrower DIBs. For example, DIB 5779 with $FWHM$ = 15.5 \AA~listed in \citetalias{Sonnentrucker2018} can be described with DIBs from 5766.30 \AA~to 5788.51 \AA, see Appendix \ref{sec:appendix_B} for a whole list of DIBs. 

We assume two possible scenarios when fitting broad absorption features: one where narrower DIB components completely describe a broad DIB, and one where some narrower DIBs are not in any way related to an overlaid broad DIB. To confirm these scenarios, we check the slope values in the $EW$ to $E(\mathrm{B-V})$ relation for DIBs whose profile resembles a broad DIB. We expect narrow DIB components that are a part of a broad DIB to have similar slopes. Below we list examples of broad features that are most protruding in the ISM spectra (Fig. \ref{fig:Figure 11}, \ref{fig:Figure 12} and \ref{fig:Figure 13}). We find that the broad features are mostly made up by DIBs with similar slopes, however in almost all cases at least one DIB has somewhat different slope than the others. This is seen in broad features at 4727 \AA, 4762 \AA, 5667 \AA, 5779 \AA~where three DIBs (including the prominent DIB 5780) all have rather different slopes, at 6495 \AA~and 6591 \AA. It is also possible that DIBs in a broad feature form more than one broad DIB, because they appear in more groups based on their slope values. This is evident in broad features at 4855 \AA, and 5740 \AA~which has at least three groups of DIBs. We also find an example of a broad feature at 6532 \AA~where slopes of all DIBs in the broad feature differ considerably, indicating that these are very likely separate and unrelated DIBs. Using multiple narrower DIBs in order to describe broad DIBs thus seems reasonable, however no reliable conclusions about the widths and components of broad DIBs can be given. It is important to note that all DIBs in our survey must satisfy the condition that their depth increases with increasing reddening and that their relation of equivalent width against the reddening can be fitted well with free parameters, such as the slope and the intercept.

Some of the DIBs are quite narrow, six of them have $FWHM <$ 0.5 \AA~and the narrowest has $FWHM \simeq 0.35$ \AA. These DIBs are also very weak, with $EW <$ 7.08 m\AA~and they all reside in the red band. Of the 6 DIBs, two (DIB 6552.47 and DIB 6574.08) were fitted manually due to their narrow width which can in some cases cause issues during the fitting procedure, i.e. fitting a broader and shallower band while excluding the narrow feature. This issue generally occurs when the narrow and the broad feature both have minima at a similar central wavelength. The fitting model does not recognize the narrow feature and fits through the broader one. One would have to use a model of the sum of two Gaussians to fit the combined profile effectively.  

Somewhat broader DIBs, 29 in total, with $FWHM$s between 0.5~\AA~and 1~\AA, are quite equally represented in all three bands. Their $EW$s span from 2.16 m\AA~to 119 m\AA, the latter belongs to DIB 5797.19.  

We report the detection of one atomic ISM absorption line, that is Li I at the central wavelength of 6707.93 \AA, already reported in \citetalias{Tuairisg2000} and \citetalias{Weselak2000}. Its strength depends on the LoS of the observed star, and is reportedly detectable in $\zeta$-type clouds only \citepalias{Weselak2000}. \citetalias{Tuairisg2000} measured its strength in three different LoSs, with $EW <$ 13 m\AA, the strongest line at the reddening of 1.01 mag. This is similar to our measured $EW$, which is 7.09 m\AA~$\pm$ 2.78 m\AA~at the reddening of 1 mag.   

We do not fit 51 DIBs mentioned in the literature (see Table \ref{tab:table_dibs_notreported}) where many of them are listed among the DIB candidates. The reason for that is a manual selection of DIB candidates in our fitting procedure, where our goal is to use the least number of DIBs possible to fit the averaged ISM spectra. Thus, some DIB candidates are left out. 11 DIBs from the literature were overlooked by our DIB seeking method. There could be several reasons why, first the search for peaks in the spectra uses some assumptions regarding the peak width, the distance between the neighbouring peaks and the minimal flux threshold, see Sec. \ref{Finding DIBs}. Second, some DIBs in the literature can appear blended in our case because we combine more than 1000 ISM spectra with different radial velocities per $E(\mathrm{B-V})$ bin. Third, some DIBs reported in the literature can only be observed in spectra of specific LoSs, while our survey provides averaged ISM spectra built from a number of different LoSs.      

We present some statistics of the 31 newly discovered DIBs. As mentioned above, 17 are categorized as certain and 14 as probable DIBs based on the relative uncertainty of their parameters, see Table \ref{tab:table_criterion}. We detected 12 new DIBs in the blue band, 9 in the green band and 10 in the red band. The newly discovered DIBs do not deviate from the others in terms of their properties. Parameters of broad newly discovered DIBs should be used with care, as we cannot exclude the possibility that they are a blend of multiple narrow DIBs.

Typically it is assumed that the $EW$ is at least roughly proportional to reddening so one would expect that the fitted value of the $EW$ at zero reddening was zero (within the measurement errors). The catalogue (Appendix \ref{sec:appendix_B}) shows that this is generally not the case: 59 out of 89 DIBs listed as certain have an equivalent width at zero reddening greater than zero at $\geq 3\,\sigma$ level. Such a significant excess of the $EW$ of a DIB implies that the absorption of the interstellar gas can happen independently of the interstellar dust absorption. We explain this by different clumpiness of the dust clouds and molecular (DIB carriers) clouds. The ISM in the Galaxy is known to be in-homogeneous at all spatial scales \citep{Ostrovskii2020}. Dust clouds have smaller spatial correlation scales than molecular clouds. Therefore, LoSs with more DIB absorption than dust absorption are more common than other way around. Hence, we can detect DIBs and measure their $EW$s even if they lie in the LoSs toward stars that are barely reddened. A significant excess can alternatively be explained if our LoSs lie toward targets that are located behind dense cloud cores. We expect to see a decrease in the ratio of the $EW$ to reddening, which becomes more apparent at higher values of reddening due to self-shielding. The spectra that contribute to a higher reddening, e.g. $E(\mathrm{B-V})$ = 0.65 mag, could correspond to a larger fraction of LoSs crossing the dense parts of the intervening clouds, leading to a lower $EW$ to reddening ratio. Extrapolating the linear relationship of the $EW$ and the reddening could consequently give a positive equivalent width at zero reddening. This phenomenon is reported in \citet{Elyajouri2017, Elyajouri2019} for DIB 15,273 observed in spectra of stars located behind the dense cloud cores. However, it is more likely that we mainly observe targets that are not situated behind dense cloud cores but are in fact far enough that the reddening is high due to more dust accumulated along the LoS.

We notice that the offset of the $EW$ also varies between different DIBs. As the LoSs are the same for our DIBs, we assume that the difference in offset is caused by the clumpiness of the DIB-bearing clouds, with variations of correlation scales in the order of $\sim$pc \citep{Points2004, VanLoon2009}. \citet{Kos2017} shows that the angular correlations of column densities of different DIBs, interstellar atomic and molecular lines vary significantly from DIB to DIB and suggests that there is a great diversity of clumpiness of the DIB-bearing clouds which occupy the same general space. Angular correlation scales from \citet{Kos2017} translate to spatial scales of about one pc.

Our LoSs also traverse $\sim$100 pc wide Local Bubble or Local Cavity around the Sun \citep{Cox1998}. This region is devoid of dust, and is filled with hot and ionized gas \citep{Lallement2015}. These conditions are probably too harsh for DIB carrier molecules to withstand, thus we are unlikely to find DIBs in the Local Cavity. Nevertheless, the diffuse molecular clouds outside of the Local Cavity become apparent closer to its border than the dust clouds. This condition assures that the slope in the $EW$ to $E(\mathrm{B-V})$ relation is always positive at low reddening as well. 

\section{Conclusions} \label{conclusion}
We divided more than 872,000 mid-to-high resolution spectra from the GALAH survey in two parts (target spectra and low-reddened spectra), based on their reddening obtained from the SFD catalogue \citep{SFD2011}. We obtained the ISM absorption spectra from the ratio of the target spectra with their low-reddened counterparts combined to the template spectra. Next we shifted the ISM absorption spectra to a common reference frame using the radial velocity shift of the DIB 5780. We were able to construct averaged ISM spectra in 6 reddening bins between $E(\mathrm{B-V})$ = 0.15 and 0.65 mag based on reddening values measured by \citet{GaiaDR3-2022,Gaia2016Prusti}. The main result is the first detailed catalogue of DIBs in the GALAH survey. It covers a total of 640 \AA~across three wavelength ranges: 4718--4889, 5653--5868, and 6481--6735 \AA. The results can be summarized as follows:

\begin{enumerate}
\item We detect a total of 183 DIBs. We present their central wavelengths, the widths and asymmetry indices as well as the slopes and the intercepts of the regression line in the relation of $EW$ against $E(\mathrm{B-V})$ in Appendix \ref{sec:appendix_B}. 31 (17$\%$) DIBs were not reported in seven modern surveys of DIBs which are based on high resolution (except for \citetalias{Sonnentrucker2018}) spectra of hot stars. We find that the majority (65$\%$) of these DIBs are broader than the median $FWHM$ for all DIBs, i.e. > 1.87 \AA. Newly discovered DIBs are almost equally represented between the spectral bands. We identify 119 DIBs that are already reported in the literature. 
\item We cross-reference our measured central wavelengths and widths with the ones noted in other DIB catalogues and find our results to be consistent with them. This excludes some broader DIBs where our measured widths tend to be broader than reported in the literature. This may be a consequence of our adopted measurement procedure which assumes that each DIB can be fitted with a single (and possibly skewed) Gaussian profile.

\item We find that the broad DIBs can be fitted with a multitude of narrower DIBs. Their widths however are not completely reliable and they depend on the DIBs selection method. Nevertheless, the width of the envelope of multiple narrow DIBs provides a lower estimate of the width of the broad DIB.

\item We detect an atomic ISM absorption line, Li I at the central wavelength of 6707.93 \AA. We measure its strength $EW$ = 7.09 m\AA~$\pm$ 2.78 m\AA~at the reddening of 1 mag and find that it is consistent with the ones reported in the literature. 

\item We find that the interstellar gas (DIB carriers) absorption can happen independently of the interstellar dust absorption, as we observe a significant excess of the $EW$ at zero reddening for the majority of certain DIBs. We explain this with our LoSs generally traversing more uniform DIB-bearing clouds compared to clumpier dust clouds. Thus, LoSs with more DIB absorption than dust absorption are more common. Additionally, we suggest a very diverse clumpiness of the DIB-bearing molecular clouds, as the non-zero offset of the $EW$ differs from DIB to DIB. We further explain the characteristics of the DIB to reddening relation with our LoSs traversing the Local Cavity. 

\end{enumerate}

\section*{Acknowledgements}

This work made use of the Third Data Release of the GALAH Survey \citep{Buder2021}. The GALAH Survey is based on data acquired through the Australian Astronomical Observatory, under programs: A/2013B/13 (The GALAH pilot survey); A/2014A/25, A/2015A/19, A2017A/18 (The GALAH survey phase 1); A2018A/18 (Open clusters with HERMES); A2019A/1 (Hierarchical star formation in Ori OB1); A2019A/15 (The GALAH survey phase 2); A/2015B/19, A/2016A/22, A/2016B/10, A/2017B/16, A/2018B/15 (The HERMES-TESS program); and A/2015A/3, A/2015B/1, A/2015B/19, A/2016A/22, A/2016B/12, A/2017A/14 (The HERMES K2-follow-up program). We acknowledge the traditional owners of the land on which the AAT stands, the Gamilaraay people, and pay our respects to elders past and present. This paper includes data that has been provided by AAO Data Central (\url{datacentral.org.au}). 

RV, JK, TZ, GT, K\v{C} and KLB thank financial support of the Slovenian Research Agency (research core funding No. P1-0188)

This work has made use of data from the European Space Agency (ESA) mission
{\it Gaia} (\url{https://www.cosmos.esa.int/gaia}), processed by the {\it Gaia}
Data Processing and Analysis Consortium (DPAC,
\url{https://www.cosmos.esa.int/web/gaia/dpac/consortium}). Funding for the DPAC
has been provided by national institutions, in particular the institutions
participating in the {\it Gaia} Multilateral Agreement.

This work was supported by the Australian Research Council Centre of Excellence for All Sky Astrophysics in 3 Dimensions (ASTRO 3D), through project number CE170100013. SLM acknowledges the support of the Australian Research Council through Discovery Project grant DP180101791, and the support of the UNSW Scientia Fellowship Program.

\section*{Data Availability}

The averaged ISM spectra and the synthetic ISM spectra shown in Fig. \ref{fig:Figure 11}, \ref{fig:Figure 12} and \ref{fig:Figure 13} along with the complete table of DIBs (see Appendix \ref{sec:appendix_B}) are available in a standardised electronic form at the CDS.



\bibliographystyle{mnras}
\bibliography{bibliography} 




\appendix

\section{Finding the DIB candidates}

\label{sec:appendix_A}

The following figures show averaged ISM spectra with different amount of reddening, their mutual correlation plot, the slope and the intercept plot of the regression line in the $E(\mathrm{B-V})$ to flux relation. DIB candidates are obtained by finding the peaks in the shown plots, where peak distance and width criterion must be met, see Sec. \ref{Finding DIBs}.
\begin{landscape}
\begin{figure}
\includegraphics[scale=0.94]{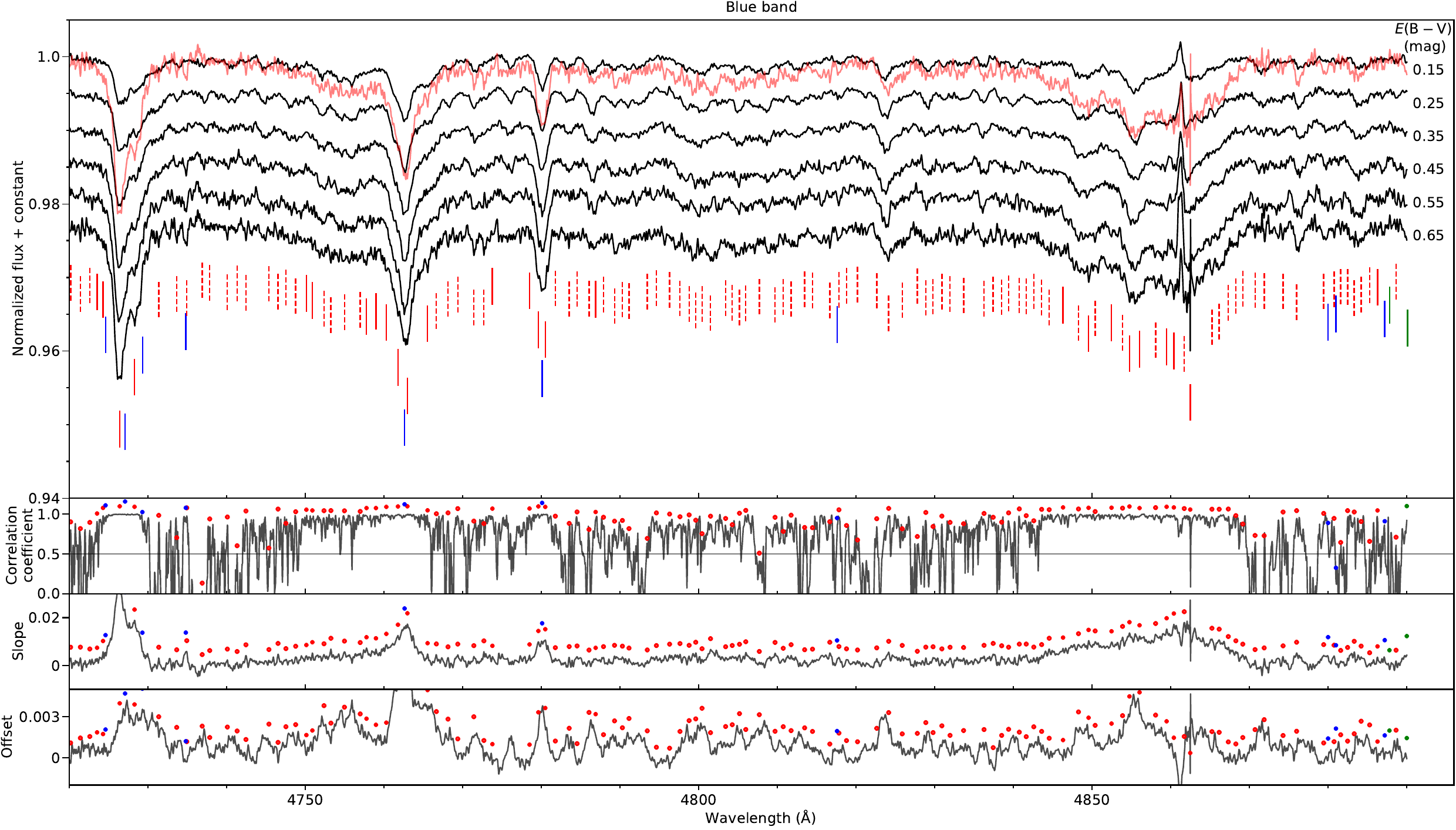}
\caption{The top panel shows averaged ISM spectra with different mean $E\mathrm{(B-V)}$ in the blue band. The averaged ISM spectrum with $E\mathrm{(B-V)}$ = 0.65 (red line) is plotted against the averaged ISM spectrum with $E\mathrm{(B-V)}$ = 0.15 in order to reveal DIBs that are hardly visible at lower reddening. Red vertical lines indicate central wavelengths of DIB candidates, while blue vertical lines indicate known central wavelengths of DIBs from catalogues \citetalias{Jenniskens1994, HobbsYork2009}. Vertical red dashed lines indicate that the DIB candidate does not have strictly successively increasing depth with the increasing amount of reddening. Green vertical lines indicate central positions of known atoms and molecules in the ISM. Red dots indicate locations where peaks from the slope plot match with either the correlation coefficient plot or the offset plot. Their positions coincide with vertical red lines in the top panel. The blue and the green dots coincide with vertical blue and green lines, also illustrated in the top panel.}
\label{fig:Figure 6}
\end{figure}
\end{landscape}

\begin{landscape}
\begin{figure}
\includegraphics[scale=0.94]{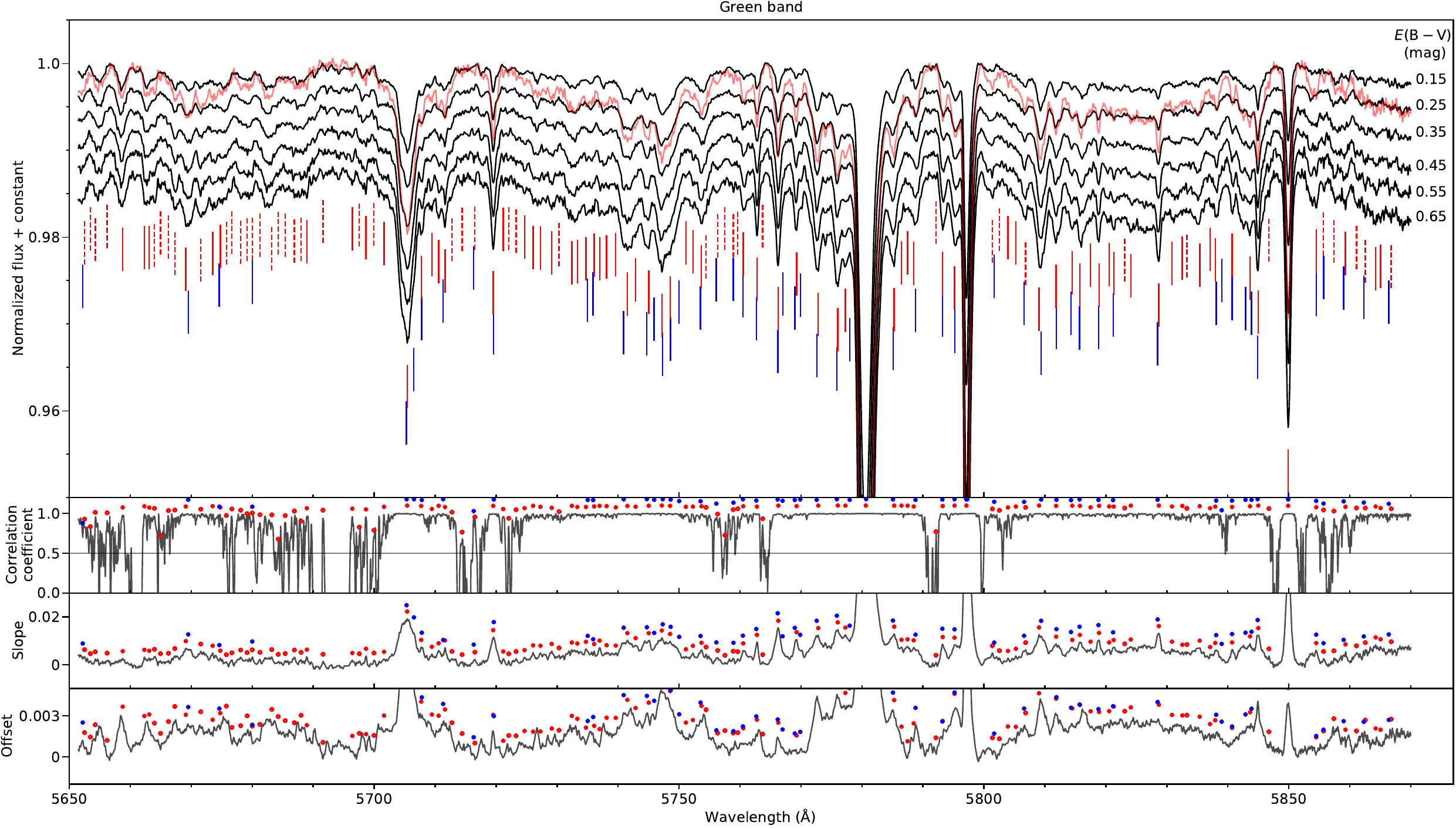}
\caption{Same as \ref{fig:Figure 6}, but for the green band.}
\label{fig:Figure 7}
\end{figure}
\end{landscape}

\begin{landscape}
\begin{figure}
\includegraphics[scale=0.94]{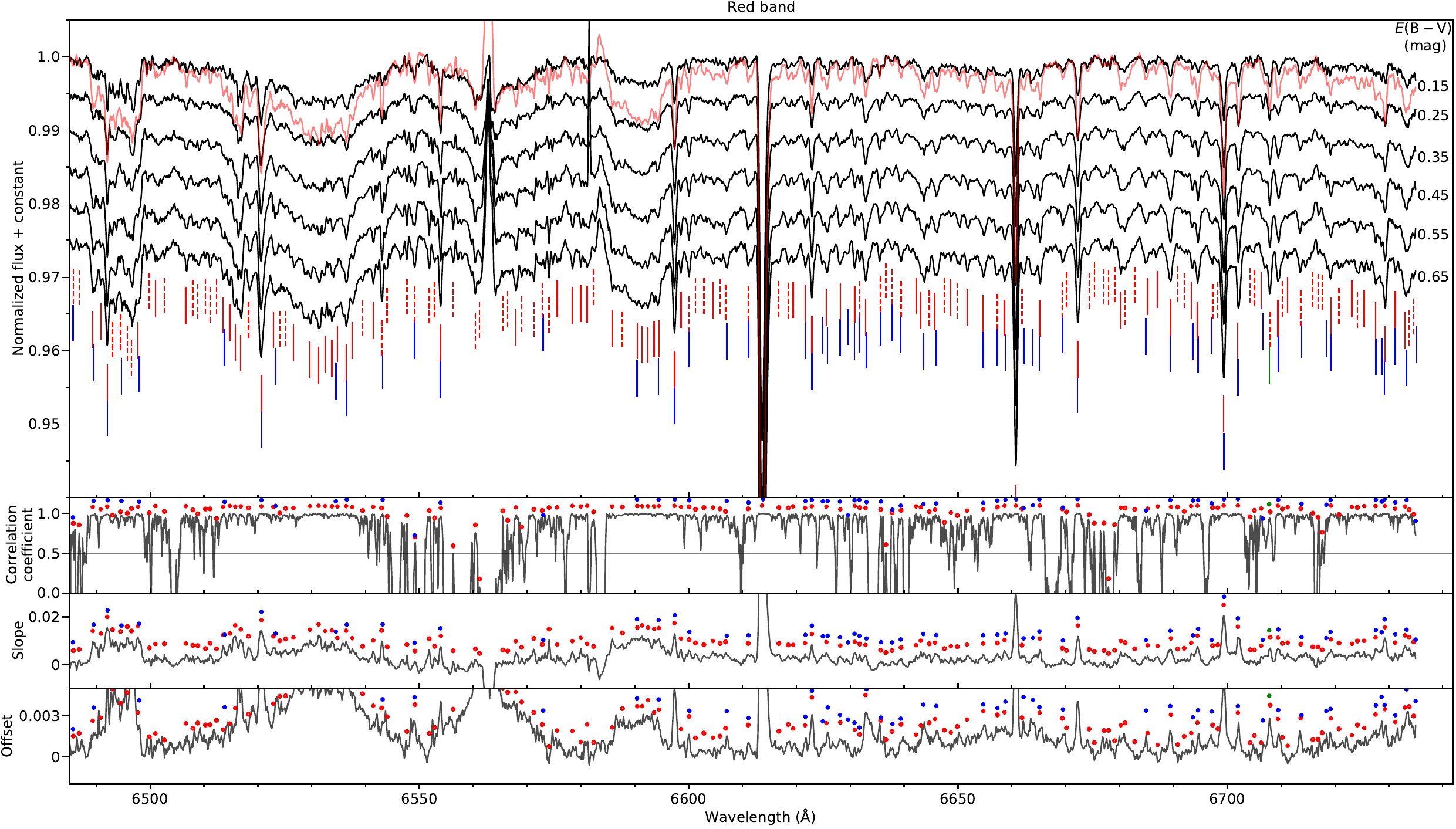}
\caption{Same as \ref{fig:Figure 6}, but for the red band.}
\label{fig:Figure 8}
\end{figure}
\end{landscape}

\section{The tables of the DIB parameters}
\label{sec:appendix_B}

We present a complete DIB catalogue of the GALAH survey that includes DIB's quality class, parameters such as the central wavelength, $FWHM$ and asymmetry indices of the Gaussian fit, the slope and the intercept of the regression line in the relation of DIB's $EW$ against the $E(\mathrm{B-V})$. To check the validity of DIB's central wavelength and width we do a cross-reference with DIBs reported in seven modern surveys based on high resolution spectra (except for \citetalias{Sonnentrucker2018}) of hot stars. The catalogue includes a total of 183 DIBs, 31 are newly discovered DIBs and 8 have a cross-reference with DIBs noted in \citetalias{HobbsYork2009} (Table 3) where they are defined as possible DIBs.
\begin{landscape}
\begin{table}
\begin{center}
\centering
\caption{The quality (Q) of each DIB is divided in three classes based on the relative uncertainty of its parameters: certain (\texttt{+}), probable (\texttt{$\circ$}) and possible (\texttt{-}). For each DIB, one can find the value of the central wavelength ($\mathrm{\lambda_C}$), the width ($FWHM$), the asymmetry ($\gamma$) of the Gaussian fit of the DIB where available, the slope and the intercept of the regression line in the relation of equivalent width against the reddening. $\lambda_{\mathrm{REF}}$ ($FWHM_{\mathrm{REF}}$) shows cross-reference values for DIBs based on their central wavelength and if available their $FWHM$. The comment denoted with ($\dagger$) marks the DIBs that were fitted manually, and ($\ddag$) marks new DIBs that have no correspondence in the reference DIB catalogues.}
\begin{threeparttable}
\label{tab:table_dibs_1}
\begin{tabular}{l c c c c c c l}
\hline
Q & $\lambda_{C}$ & $\lambda_{\mathrm{REF}}$ ($FWHM_{\mathrm{REF}}$) & $FWHM$ & $\gamma$ & $\Delta EW$ / $\Delta E(\mathrm{B-V})$ & $EW(E\mathrm{(B-V) = 0})$ & Comment\\
& (\AA) & (\AA) & (\AA) & & ($\mathrm{m}$\AA\ / $\mathrm{mag}$) & ($\mathrm{m}$\AA) & \\\hline\hline
\texttt{-} & 4722.54 $\pm$ 0.75 & & 3.92 $\pm$ 1.45 & & 8.85 $\pm$ 1.81 & 0.94 $\pm$ 0.78 & \\
\texttt{+} & 4726.40 $\pm$ 0.07 & 4727.16 (3.11)$^{\mathrm{a}}$, 4726.27$^{\mathrm{b}}$, 4726.35 (1.70)$^{\mathrm{c}}$, 4726.59 (1.25)$^{\mathrm{e}}$, 4726.98 (2.75)$^{\mathrm{f}}$ & 1.39 $\pm$ 0.07 & & 45.89 $\pm$ 1.25 & 1.15 $\pm$ 0.54 & $\dagger$\\
\texttt{+} & 4728.28 $\pm$ 0.12 & 4728.35 (1.54)$^{\mathrm{c}}$ & 1.87 $\pm$ 0.12 & & 38.57 $\pm$ 1.89 & 2.49 $\pm$ 0.82 & $\dagger$\\
\texttt{-} & 4731.60 $\pm$ 0.43 & & 4.47 $\pm$ 0.18 & & 8.12 $\pm$ 4.12 & 5.16 $\pm$ 1.79 & \\
\texttt{-} & 4745.87 $\pm$ 1.09 & & 6.83 $\pm$ 0.94 & & 7.80 $\pm$ 3.13 & 5.38 $\pm$ 1.36 & \\
\texttt{$\circ$} & 4751.87 $\pm$ 0.31 & & 4.68 $\pm$ 0.59 & & 17.66 $\pm$ 1.46 & 6.91 $\pm$ 0.63 & $\ddag$\\
\texttt{-} & 4754.58 $\pm$ 0.22 & & 1.87 $\pm$ 0.43 & & 6.52 $\pm$ 2.36 & 2.35 $\pm$ 1.02 & \\
\texttt{-} & 4755.98 $\pm$ 0.13 & & 1.77 $\pm$ 0.25 & & 8.27 $\pm$ 6.07 & 3.38 $\pm$ 2.63 & \\
\texttt{$\circ$} & 4758.19 $\pm$ 0.23 & 4757.89 (14.45)$^{\mathrm{a\ast}}$ & 3.77 $\pm$ 0.44 & & 22.08 $\pm$ 3.73 & 3.63 $\pm$ 1.62 & \\
\texttt{+} & 4762.57 $\pm$ 0.06 & 4762.62 (2.50)$^{\mathrm{a}}$, 4762.67$^{\mathrm{b}}$, 4762.57 (2.89)$^{\mathrm{c}}$, 4762.57 (2.10)$^{\mathrm{e}}$, 4762.44 (1.94)$^{\mathrm{f}}$, 4761+4764 (25.33)$^{\mathrm{g}}$ & 3.54 $\pm$ 0.11 & & 60.30 $\pm$ 3.04 & 20.50 $\pm$ 1.32 & \\
\texttt{-} & 4766.15 $\pm$ 0.14 & & 2.57 $\pm$ 0.26 & & 3.76 $\pm$ 2.44 & 8.89 $\pm$ 1.06 & \\
\texttt{-} & 4768.43 $\pm$ 0.33 & & 2.33 $\pm$ 0.64 & & 5.44 $\pm$ 4.72 & 2.00 $\pm$ 2.04 & \\
\texttt{-} & 4771.41 $\pm$ 0.36 & & 0.60 $\pm$ 0.36 & & 1.44 $\pm$ 0.79 & 1.12 $\pm$ 0.34 & $\dagger$\\
\texttt{-} & 4772.69 $\pm$ 0.28 & & 0.57 $\pm$ 0.28 & & 3.15 $\pm$ 0.85 & -0.14 $\pm$ 0.37 & $\dagger$\\
\texttt{+} & 4780.10 $\pm$ 0.08 & 4780.24 (1.72)$^{\mathrm{a}}$, 4780.04$^{\mathrm{b}}$, 4780.10 (1.74)$^{\mathrm{c}}$, 4780.09 (1.45)$^{\mathrm{e}}$, 4780.10 (1.54)$^{\mathrm{f}}$ & 1.52 $\pm$ 0.15 & & 21.12 $\pm$ 0.89 & 3.10 $\pm$ 0.39 & \\
\texttt{-} & 4783.46 $\pm$ 0.35 & & 1.95 $\pm$ 0.68 & & 8.18 $\pm$ 3.24 & 0.45 $\pm$ 1.40 & \\
\texttt{-} & 4786.49 $\pm$ 0.17 & & 1.57 $\pm$ 0.33 & & 3.67 $\pm$ 2.99 & 3.38 $\pm$ 1.29 & \\
\texttt{-} & 4788.97 $\pm$ 0.40 & & 2.11 $\pm$ 0.78 & & 3.99 $\pm$ 1.91 & 1.74 $\pm$ 0.83 & \\
\texttt{$\circ$} & 4791.58 $\pm$ 0.40 & & 4.04 $\pm$ 0.79 & & 12.91 $\pm$ 4.81 & 3.66 $\pm$ 2.08 & $\ddag$\\
\texttt{+} & 4799.92 $\pm$ 0.25 & & 4.76 $\pm$ 0.56 & & 23.50 $\pm$ 5.17 & 6.54 $\pm$ 2.24 & $\ddag$\\
\texttt{$\circ$} & 4805.68 $\pm$ 0.77 & & 4.68 $\pm$ 0.62 & & 19.94 $\pm$ 4.88 & 5.93 $\pm$ 2.12 & $\ddag$\\
\texttt{-} & 4809.20 $\pm$ 0.19 & & 2.96 $\pm$ 0.64 & & 9.02 $\pm$ 2.98 & 3.22 $\pm$ 1.29 & \\
\texttt{-} & 4812.24 $\pm$ 0.32 & & 1.48 $\pm$ 0.62 & & 6.16 $\pm$ 2.13 & 0.30 $\pm$ 0.92 & \\
\texttt{-} & 4814.65 $\pm$ 0.60 & & 3.31 $\pm$ 1.16 & & 4.76 $\pm$ 0.86 & 2.51 $\pm$ 0.37 & \\
\texttt{-} & 4817.55 $\pm$ 0.93 & 4817.62 (0.35)$^{\mathrm{f}}$ & 3.46 $\pm$ 1.76 & & 14.95 $\pm$ 5.52 & -1.90 $\pm$ 2.39 & \\
\texttt{$\circ$} & 4823.72 $\pm$ 0.15 & 4824.00 (32.47)$^{\mathrm{e}}$ & 2.06 $\pm$ 0.30 & & 9.11 $\pm$ 4.10 & 4.91 $\pm$ 1.78 & \\
\texttt{-} & 4825.85 $\pm$ 0.44 & 4825.91 (20.40)$^{\mathrm{c}}$ & 1.71 $\pm$ 0.84 & & 1.92 $\pm$ 1.33 & 1.46 $\pm$ 0.58 & \\
\texttt{$\circ$} & 4829.30 $\pm$ 0.41 & & 2.89 $\pm$ 0.81 & & 14.86 $\pm$ 5.40 & 0.09 $\pm$ 2.34 & $\ddag$\\
\texttt{-} & 4833.16 $\pm$ 0.08 & & 3.88 $\pm$ 1.19 & & 8.74 $\pm$ 2.42 & 2.30 $\pm$ 1.05 & \\
\texttt{-} & 4836.19 $\pm$ 0.24 & & 0.90 $\pm$ 0.46 & & 4.27 $\pm$ 0.88 & 0.11 $\pm$ 0.38 & \\
\texttt{-} & 4838.51 $\pm$ 0.45 & & 0.53 $\pm$ 0.45 & & 1.54 $\pm$ 0.68 & 0.14 $\pm$ 0.29 & $\dagger$\\
\texttt{-} & 4839.89 $\pm$ 0.24 & & 1.90 $\pm$ 0.61 & & 7.02 $\pm$ 2.11 & 1.16 $\pm$ 0.92 & \\
\texttt{$\circ$} & 4843.49 $\pm$ 0.38 & & 3.49 $\pm$ 0.74 & & 13.95 $\pm$ 2.20 & 2.02 $\pm$ 0.95 & $\ddag$\\
\texttt{+} & 4848.77 $\pm$ 0.14 & & 4.46 $\pm$ 0.33 & & 46.05 $\pm$ 3.41 & 5.00 $\pm$ 1.48 & $\ddag$\\
\texttt{+} & 4855.25 $\pm$ 0.12 & & 4.95 $\pm$ 0.24 & & 59.35 $\pm$ 4.67 & 14.36 $\pm$ 2.02 & $\ddag$\\
\hline
\end{tabular}
\begin{tablenotes}
      \small
      \item Cross-reference with a DIB noted in \texttt{a}: \citetalias{HobbsYork2009}; \texttt{b}: \citetalias{GalazutdinovMusaev2000}; \texttt{c}: \citetalias{Tuairisg2000}; \texttt{e}: \citetalias{Jenniskens1994}; \texttt{f}: \citetalias{Fan2019}; \texttt{g}: \citetalias{Sonnentrucker2018};
      \texttt{a$\ast$}: Cross-reference with a DIB noted in \citetalias{HobbsYork2009} (Table 3) defined as a possible DIB.
    \end{tablenotes}
\end{threeparttable}
\end{center}
\end{table}
\end{landscape}

\begin{landscape}
\begin{table}
\centering
\caption{continued.}
\begin{threeparttable}
\label{tab:table_dibs_2}
\begin{tabular}{l c c c c c c l}
\hline
Q & $\lambda_{C}$ & $\lambda_{\mathrm{REF}}$ ($FWHM_{\mathrm{REF}}$) & $FWHM$ & $\gamma$ & $\Delta EW$ / $\Delta E(\mathrm{B-V})$ & $EW(E\mathrm{(B-V) = 0})$ & Comment\\
& (\AA) & (\AA) & (\AA) & & ($\mathrm{m}$\AA\ / $\mathrm{mag}$) & ($\mathrm{m}$\AA) & \\\hline\hline
\texttt{+} & 4859.89 $\pm$ 0.01 & & 4.28 $\pm$ 0.38 & & 56.75 $\pm$ 7.11 & -0.35 $\pm$ 3.08 & $\ddag$\\
\texttt{+} & 4862.81 $\pm$ 0.14 & & 1.72 $\pm$ 0.26 & & 19.07 $\pm$ 1.44 & 0.79 $\pm$ 0.63 & $\ddag$\\
\texttt{$\circ$} & 4864.73 $\pm$ 0.24 & & 2.54 $\pm$ 0.48 & & 14.32 $\pm$ 2.10 & 2.28 $\pm$ 0.91 & $\ddag$\\
\texttt{+} & 4866.65 $\pm$ 0.07 & & 4.47 $\pm$ 0.11 & & 31.30 $\pm$ 7.82 & -2.50 $\pm$ 3.39 & $\ddag$\\
\texttt{-} & 4876.21 $\pm$ 0.38 & & 0.90 $\pm$ 0.38 & & 3.50 $\pm$ 0.79 & 0.96 $\pm$ 0.34 & $\dagger$\\
\texttt{-} & 4879.93 $\pm$ 0.33 & 4880.12 (1.32)$^{\mathrm{a\ast}}$, 4880.35 (1.35)$^{\mathrm{e}}$ & 2.23 $\pm$ 1.56 & & 5.85 $\pm$ 2.53 & -0.14 $\pm$ 1.10 & \\
\texttt{-} & 4884.01 $\pm$ 0.31 & & 2.17 $\pm$ 0.59 & & 8.14 $\pm$ 3.43 & 1.98 $\pm$ 1.49 & \\
\texttt{-} & 4886.56 $\pm$ 0.89 & 4887.17 (0.85)$^{\mathrm{f}}$ & 2.59 $\pm$ 2.45 & 1.11 $\pm$ 2.83 & 4.55 $\pm$ 4.52 & 0.35 $\pm$ 1.96 & \\
\texttt{-} & 4888.72 $\pm$ 0.45 & & 0.69 $\pm$ 0.86 & & -0.80 $\pm$ 1.57 & 0.83 $\pm$ 0.68 & \\
\texttt{-} & 5654.75 $\pm$ 0.21 & & 1.87 $\pm$ 1.30 & & 1.64 $\pm$ 1.77 & 4.11 $\pm$ 0.77 & \\
\texttt{$\circ$} & 5658.61 $\pm$ 0.44 & & 1.47 $\pm$ 0.85 & & 3.29 $\pm$ 0.36 & 3.96 $\pm$ 0.15 & $\ddag$\\
\texttt{$\circ$} & 5662.80 $\pm$ 2.02 & & 2.82 $\pm$ 1.43 & -0.14 $\pm$ 0.85 & 3.30 $\pm$ 0.60 & 6.86 $\pm$ 0.26 & $\ddag$\\
\texttt{$\circ$} & 5667.39 $\pm$ 0.63 & & 2.81 $\pm$ 1.20 & & 4.92 $\pm$ 0.72 & 7.50 $\pm$ 0.31 & $\ddag$\\
\texttt{-} & 5669.55 $\pm$ 0.67 & 5669.63 (1.34)$^{\mathrm{a}}$, 5669.12 (0.67)$^{\mathrm{f}}$, 5669.94 (0.53)$^{\mathrm{f}}$ & 1.80 $\pm$ 1.29 & & 7.03 $\pm$ 4.80 & 2.14 $\pm$ 2.08 & \\
\texttt{$\circ$} & 5671.97 $\pm$ 0.73 & 5671.89 (1.39)$^{\mathrm{f}}$ & 2.59 $\pm$ 1.83 & -0.76 $\pm$ 1.92 & 10.92 $\pm$ 1.80 & 2.80 $\pm$ 0.78 & \\
\texttt{-} & 5674.66 $\pm$ 0.94 & & 4.41 $\pm$ 1.86 & 0.42 $\pm$ 0.77 & 0.78 $\pm$ 4.54 & 11.07 $\pm$ 1.97 & \\
\texttt{-} & 5678.97 $\pm$ 0.95 & & 2.03 $\pm$ 1.84 & & 6.22 $\pm$ 5.26 & 1.68 $\pm$ 2.28 & \\
\texttt{-} & 5682.84 $\pm$ 0.64 & & 3.48 $\pm$ 1.24 & & 7.79 $\pm$ 3.08 & 9.74 $\pm$ 1.33 & \\
\texttt{$\circ$} & 5687.58 $\pm$ 1.10 & & 4.80 $\pm$ 2.04 & -0.15 $\pm$ 0.46 & 10.30 $\pm$ 2.57 & 8.76 $\pm$ 1.11 & $\ddag$\\
\texttt{-} & 5699.16 $\pm$ 2.41 & & 6.28 $\pm$ 4.51 & & -0.42 $\pm$ 1.37 & 8.11 $\pm$ 0.60 & \\
\texttt{+} & 5705.21 $\pm$ 0.16 & 5705.31 (2.68)$^{\mathrm{a}}$, 5705.20$^{\mathrm{b}}$, 5705.10 (2.75)$^{\mathrm{c}}$, 5705.43$^{\mathrm{d}}$, 5705.13 (2.15)$^{\mathrm{e}}$, 5705.12 (2.68)$^{\mathrm{f}}$ & 2.81 $\pm$ 0.30 & 0.27 $\pm$ 0.20 & 48.88 $\pm$ 7.84 & 21.59 $\pm$ 3.40 & \\
\texttt{+} & 5707.71 $\pm$ 0.14 & 5707.84 (0.45)$^{\mathrm{a}}$, 5707.72 (0.61)$^{\mathrm{f}}$ & 1.22 $\pm$ 0.14 & & 9.47 $\pm$ 0.40 & 3.26 $\pm$ 0.17 & $\dagger$\\
\texttt{$\circ$} & 5710.56 $\pm$ 0.20 & 5710.58 (0.58)$^{\mathrm{f}}$ & 1.12 $\pm$ 0.20 & & 6.99 $\pm$ 0.30 & 1.16 $\pm$ 0.13 & $\dagger$\\
\texttt{+} & 5711.61 $\pm$ 0.15 & 5711.17 (1.49)$^{\mathrm{a}}$, 5711.46 (1.23)$^{\mathrm{f}}$ & 1.09 $\pm$ 0.15 & & 8.00 $\pm$ 0.65 & 2.30 $\pm$ 0.28 & $\dagger$\\
\texttt{+} & 5719.56 $\pm$ 0.28 & 5719.63 (0.92)$^{\mathrm{a}}$, 5719.30$^{\mathrm{b}}$, 5719.68 (1.04)$^{\mathrm{c}}$, 5719.40$^{\mathrm{d}}$, 5719.43 (0.93)$^{\mathrm{e}}$, 5719.57 (0.80)$^{\mathrm{f}}$ & 1.09 $\pm$ 0.55 & & 14.13 $\pm$ 0.63 & 0.38 $\pm$ 0.27 & \\
\texttt{-} & 5725.17 $\pm$ 0.45 & & 5.59 $\pm$ 5.03 & 0.41 $\pm$ 1.15 & 10.56 $\pm$ 1.50 & 4.00 $\pm$ 0.65 & \\
\texttt{+} & 5729.04 $\pm$ 1.09 & & 4.41 $\pm$ 1.95 & 0.07 $\pm$ 0.48 & 17.79 $\pm$ 1.30 & 5.56 $\pm$ 0.56 & $\ddag$\\
\texttt{+} & 5732.12 $\pm$ 0.51 & 5732.67 (1.42)$^{\mathrm{f}}$ & 1.94 $\pm$ 0.99 & & 10.77 $\pm$ 1.32 & 2.95 $\pm$ 0.57 & \\
\texttt{+} & 5735.22 $\pm$ 3.72 & 5734.90 (0.54)$^{\mathrm{f}}$, 5735.89 (0.46)$^{\mathrm{f}}$ & 4.94 $\pm$ 2.05 & -0.13 $\pm$ 0.42 & 39.82 $\pm$ 6.17 & 0.35 $\pm$ 2.67 & \\
\texttt{+} & 5741.28 $\pm$ 0.47 & 5740.90 (2.31)$^{\mathrm{a}}$, 5741.10 (1.89)$^{\mathrm{f}}$ & 4.64 $\pm$ 0.98 & 0.30 $\pm$ 0.29 & 40.08 $\pm$ 0.94 & 13.39 $\pm$ 0.41 & \\
\texttt{+} & 5744.52 $\pm$ 0.31 & 5744.66 (1.08)$^{\mathrm{a}}$, 5744.70 (1.20)$^{\mathrm{f}}$ & 1.58 $\pm$ 0.80 & 0.40 $\pm$ 1.62 & 12.34 $\pm$ 0.64 & 1.99 $\pm$ 0.28 & \\
\texttt{+} & 5747.62 $\pm$ 0.35 & 5748.62 (2.11)$^{\mathrm{a}}$, 5747.81 (2.32)$^{\mathrm{e}}$, 5747.15 (1.71)$^{\mathrm{f}}$ & 4.54 $\pm$ 0.68 & & 54.53 $\pm$ 3.93 & 15.88 $\pm$ 1.70 & \\
\texttt{+} & 5753.17 $\pm$ 0.20 & 5753.51 (0.55)$^{\mathrm{f}}$ & 5.22 $\pm$ 1.54 & 0.08 $\pm$ 0.27 & 22.89 $\pm$ 1.41 & 10.99 $\pm$ 0.61 & \\
\texttt{$\circ$} & 5760.53 $\pm$ 0.20 & 5760.40$^{\mathrm{b}}$, 5760.64 (0.61)$^{\mathrm{c}}$, 5760.44$^{\mathrm{d}}$, 5760.48 (0.72)$^{\mathrm{f}}$ & 1.10 $\pm$ 0.20 & & 6.87 $\pm$ 0.37 & 0.74 $\pm$ 0.16 & $\dagger$\\
\texttt{+} & 5762.78 $\pm$ 0.10 & 5762.70 (0.61)$^{\mathrm{a}}$, 5762.70$^{\mathrm{b}}$, 5762.80 (0.95)$^{\mathrm{c}}$, 5762.69$^{\mathrm{d}}$, 5762.50 (0.66)$^{\mathrm{e}}$, 5762.68 (0.54)$^{\mathrm{f}}$ & 0.75 $\pm$ 0.10 & & 7.83 $\pm$ 0.45 & 0.72 $\pm$ 0.19 & $\dagger$\\
\texttt{+} & 5766.30 $\pm$ 0.27 & 5766.25 (0.79)$^{\mathrm{a}}$, 5766.16$^{\mathrm{b}}$, 5766.17 (1.08)$^{\mathrm{c}}$, 5766.15$^{\mathrm{d}}$, 5766.25 (0.92)$^{\mathrm{e}}$, 5766.16 (0.76)$^{\mathrm{f}}$ & 1.34 $\pm$ 0.53 & & 20.35 $\pm$ 1.36 & 0.47 $\pm$ 0.59 & $\dagger$\\
\texttt{+} & 5769.39 $\pm$ 0.76 & 5769.09$^{\mathrm{b}}$, 5769.32 (1.26)$^{\mathrm{c}}$, 5769.03$^{\mathrm{d}}$, 5769.10 (0.80)$^{\mathrm{e}}$, 5769.09 (0.60)$^{\mathrm{f}}$, 5769.91 (0.49)$^{\mathrm{f}}$ & 2.67 $\pm$ 1.46 & & 17.16 $\pm$ 0.88 & 1.47 $\pm$ 0.38 & \\
\texttt{+} & 5772.55 $\pm$ 0.14 & 5772.66 (1.23)$^{\mathrm{a}}$, 5772.60$^{\mathrm{b}}$, 5772.67 (1.45)$^{\mathrm{c}}$, 5772.53$^{\mathrm{d}}$, 5772.60 (1.03)$^{\mathrm{e}}$, 5772.63 (1.23)$^{\mathrm{f}}$ & 1.95 $\pm$ 0.27 & & 22.34 $\pm$ 0.71 & 5.54 $\pm$ 0.31 & \\
\texttt{+} & 5776.03 $\pm$ 0.10 & 5775.97 (0.91)$^{\mathrm{a}}$, 5775.78$^{\mathrm{b}}$, 5776.21 (2.00)$^{\mathrm{c}}$, 5775.75$^{\mathrm{d}}$, 5776.08 (0.85)$^{\mathrm{e}}$, 5775.91 (1.00)$^{\mathrm{f}}$ & 2.21 $\pm$ 0.10 & & 34.40 $\pm$ 0.58 & 8.58 $\pm$ 0.25 & \\
\hline
\end{tabular}
\begin{tablenotes}
      \small
      \item Cross-reference with a DIB noted in \texttt{a}: \citetalias{HobbsYork2009}; \texttt{b}: \citetalias{GalazutdinovMusaev2000}; \texttt{c}: \citetalias{Tuairisg2000}; \texttt{d}: \citetalias{Weselak2000}; \texttt{e}: \citetalias{Jenniskens1994}; \texttt{f}: \citetalias{Fan2019};
      \texttt{a$\ast$}: Cross-reference with a DIB noted in \citetalias{HobbsYork2009} (Table 3) defined as a possible DIB.
    \end{tablenotes}
\end{threeparttable}
\end{table}
\end{landscape}

\begin{landscape}
\begin{table}
\centering
\caption{continued.}
\begin{threeparttable}
\label{tab:table_dibs_3}
\begin{tabular}{l c c c c c c l}
\hline
Q & $\lambda_{C}$ & $\lambda_{\mathrm{REF}}$ ($FWHM_{\mathrm{REF}}$) & $FWHM$ & $\gamma$ & $\Delta EW$ / $\Delta E(\mathrm{B-V})$ & $EW(E\mathrm{(B-V) = 0})$ & Comment\\
& (\AA) & (\AA) & (\AA) & & ($\mathrm{m}$\AA\ / $\mathrm{mag}$) & ($\mathrm{m}$\AA) & \\\hline\hline
\texttt{+} & 5780.59 $\pm$ 0.01 & 5780.61 (2.14)$^{\mathrm{a}}$, 5780.37$^{\mathrm{b}}$, 5780.55 (2.22)$^{\mathrm{c}}$, 5780.50$^{\mathrm{d}}$, 5780.59 (2.07)$^{\mathrm{e}}$, 5780.64 (2.09)$^{\mathrm{f}}$ & 2.13 $\pm$ 0.02 & -0.18 $\pm$ 0.03 & 283.87 $\pm$ 4.13 & 73.62 $\pm$ 1.79 & \\
\texttt{+} & 5784.78 $\pm$ 0.03 & 5785.09 (0.75)$^{\mathrm{a}}$, 5785.05$^{\mathrm{b}}$, 5784.86 (0.92)$^{\mathrm{c}}$, 5785.11$^{\mathrm{d}}$, 5784.90 (1.00)$^{\mathrm{e}}$, 5785.00 (0.98)$^{\mathrm{f}}$ & 4.60 $\pm$ 0.52 & 0.33 $\pm$ 0.16 & 50.60 $\pm$ 1.22 & 10.80 $\pm$ 0.53 & \\
\texttt{+} & 5788.51 $\pm$ 0.24 & 5788.76 (1.76)$^{\mathrm{a}}$, 5789.04 (1.16)$^{\mathrm{c}}$, 5788.90$^{\mathrm{d}}$, 5789.06 (1.13)$^{\mathrm{e}}$, 5788.70 (2.38)$^{\mathrm{f}}$ & 1.84 $\pm$ 0.45 & & 16.17 $\pm$ 0.51 & 1.54 $\pm$ 0.22 & \\
\texttt{-} & 5793.27 $\pm$ 0.30 & 5793.17 (0.87)$^{\mathrm{a}}$, 5793.22$^{\mathrm{b}}$, 5793.13 (0.95)$^{\mathrm{c}}$, 5793.19$^{\mathrm{d}}$, 5793.24 (0.96)$^{\mathrm{f}}$ & 1.05 $\pm$ 0.57 & & 2.25 $\pm$ 1.42 & 3.32 $\pm$ 0.61 & \\
\texttt{+} & 5795.58 $\pm$ 0.34 & 5795.21 (1.12)$^{\mathrm{a}}$, 5795.16$^{\mathrm{b}}$, 5795.20$^{\mathrm{d}}$, 5795.23 (4.09)$^{\mathrm{e}}$, 5795.21 (1.42)$^{\mathrm{f}}$ & 2.56 $\pm$ 0.61 & -0.09 $\pm$ 0.44 & 23.16 $\pm$ 2.96 & 7.88 $\pm$ 1.28 & \\
\texttt{+} & 5797.19 $\pm$ 0.03 & 5797.20 (0.91)$^{\mathrm{a}}$, 5796.96$^{\mathrm{b}}$, 5797.08 (0.96)$^{\mathrm{c}}$, 5796.97$^{\mathrm{d}}$, 5797.11 (0.97)$^{\mathrm{e}}$, 5797.18 (0.89)$^{\mathrm{f}}$ & 0.99 $\pm$ 0.05 & -0.77 $\pm$ 0.28 & 109.58 $\pm$ 4.93 & 9.76 $\pm$ 2.14 & \\
\texttt{$\circ$} & 5802.38 $\pm$ 0.22 & 5801.73 (1.46)$^{\mathrm{a}}$, 5801.47 (2.24)$^{\mathrm{f}}$ & 4.89 $\pm$ 3.80 & & 12.30 $\pm$ 1.91 & 2.60 $\pm$ 0.83 & \\
\texttt{+} & 5806.56 $\pm$ 0.71 & 5806.76$^{\mathrm{b}}$, 5806.52 (1.54)$^{\mathrm{c}}$, 5806.60 (0.45)$^{\mathrm{f}}$ & 3.97 $\pm$ 1.37 & & 17.51 $\pm$ 1.75 & 7.57 $\pm$ 0.76 & \\
\texttt{+} & 5809.42 $\pm$ 0.25 & 5809.53 (1.37)$^{\mathrm{a}}$, 5809.24$^{\mathrm{b}}$, 5809.53 (1.83)$^{\mathrm{c}}$, 5809.22$^{\mathrm{d}}$, 5809.12 (1.10)$^{\mathrm{e}}$, 5809.25 (1.17)$^{\mathrm{f}}$ & 2.04 $\pm$ 0.48 & & 35.71 $\pm$ 1.79 & 3.40 $\pm$ 0.77 & \\
\texttt{+} & 5811.87 $\pm$ 0.37 & 5811.91 (1.06)$^{\mathrm{a}}$, 5811.96$^{\mathrm{b}}$, 5811.57$^{\mathrm{d}}$, 5811.96 (1.41)$^{\mathrm{f}}$ & 2.08 $\pm$ 0.71 & & 14.68 $\pm$ 1.18 & 5.49 $\pm$ 0.51 & \\
\texttt{+} & 5814.49 $\pm$ 0.16 & 5814.21$^{\mathrm{d}}$, 5814.50 (0.40)$^{\mathrm{e}}$, 5814.28 (0.50)$^{\mathrm{f}}$ & 1.37 $\pm$ 0.16 & & 10.75 $\pm$ 0.37 & 2.85 $\pm$ 0.16 & $\dagger$\\
\texttt{+} & 5815.74 $\pm$ 0.13 & 5815.72 (0.85)$^{\mathrm{a}}$, 5815.71$^{\mathrm{b}}$, 5815.80 (0.65)$^{\mathrm{c}}$, 5815.78$^{\mathrm{d}}$, 5815.72 (0.67)$^{\mathrm{f}}$ & 1.37 $\pm$ 0.13 & & 13.87 $\pm$ 0.59 & 3.11 $\pm$ 0.25 & $\dagger$\\
\texttt{+} & 5818.89 $\pm$ 0.11 & 5818.81 (0.65)$^{\mathrm{a}}$, 5818.75$^{\mathrm{b}}$, 5818.47 (0.81)$^{\mathrm{c}}$, 5818.69$^{\mathrm{d}}$, 5818.85 (0.49)$^{\mathrm{e}}$, 5818.74 (0.54)$^{\mathrm{f}}$ & 1.01 $\pm$ 0.11 & & 11.03 $\pm$ 0.43 & 1.66 $\pm$ 0.19 & $\dagger$\\
\texttt{$\circ$} & 5821.29 $\pm$ 0.12 & 5821.26 (0.45)$^{\mathrm{a}}$, 5821.23$^{\mathrm{b}}$, 5821.22$^{\mathrm{d}}$, 5821.15 (0.43)$^{\mathrm{f}}$ & 0.84 $\pm$ 0.12 & & 6.44 $\pm$ 0.51 & 1.87 $\pm$ 0.22 & $\dagger$\\
\texttt{+} & 5824.25 $\pm$ 0.53 & & 3.23 $\pm$ 1.03 & & 14.52 $\pm$ 1.34 & 9.03 $\pm$ 0.58 & $\ddag$\\
\texttt{+} & 5826.89 $\pm$ 0.46 & & 2.46 $\pm$ 0.88 & & 19.70 $\pm$ 0.49 & 4.34 $\pm$ 0.21 & $\ddag$\\
\texttt{+} & 5828.69 $\pm$ 0.14 & 5828.63 (0.88)$^{\mathrm{a}}$, 5828.46$^{\mathrm{b}}$, 5828.56 (0.87)$^{\mathrm{c}}$, 5828.52$^{\mathrm{d}}$, 5828.40 (0.31)$^{\mathrm{e}}$, 5828.50 (0.78)$^{\mathrm{f}}$ & 1.15 $\pm$ 0.14 & & 9.17 $\pm$ 0.67 & 2.51 $\pm$ 0.29 & $\dagger$\\
\texttt{+} & 5830.84 $\pm$ 0.17 & 5830.77$^{\mathrm{d}}$ & 1.30 $\pm$ 0.17 & & 9.56 $\pm$ 0.69 & 2.08 $\pm$ 0.30 & $\dagger$\\
\texttt{+} & 5835.39 $\pm$ 0.20 & 5834.54 (0.84)$^{\mathrm{c}}$, 5835.00$^{\mathrm{d}}$ & 1.80 $\pm$ 0.20 & & 13.50 $\pm$ 0.86 & 2.63 $\pm$ 0.37 & $\dagger$\\
\texttt{$\circ$} & 5838.11 $\pm$ 0.59 & 5838.09 (0.46)$^{\mathrm{a}}$, 5838.00$^{\mathrm{b}}$, 5837.92 (0.52)$^{\mathrm{c}}$, 5838.08$^{\mathrm{d}}$, 5838.04 (0.53)$^{\mathrm{f}}$, 5839.00 (0.46)$^{\mathrm{f}}$ & 1.93 $\pm$ 1.15 & & 8.42 $\pm$ 2.07 & 3.05 $\pm$ 0.90 & \\
\texttt{-} & 5840.69 $\pm$ 0.31 & 5840.69 (0.42)$^{\mathrm{a}}$, 5840.65$^{\mathrm{b}}$, 5840.62 (0.64)$^{\mathrm{c}}$, 5840.72$^{\mathrm{d}}$, 5840.65 (0.49)$^{\mathrm{f}}$ & 1.02 $\pm$ 0.31 & & 4.79 $\pm$ 0.48 & -0.29 $\pm$ 0.21 & $\dagger$\\
\texttt{+} & 5843.54 $\pm$ 0.59 & 5843.92 (4.38)$^{\mathrm{a}}$, 5844.28 (4.61)$^{\mathrm{c}}$, 5844.15$^{\mathrm{d}}$, 5844.19 (3.04)$^{\mathrm{e}}$, 5843.50 (0.85)$^{\mathrm{f}}$ & 3.97 $\pm$ 1.13 & & 38.00 $\pm$ 9.02 & 4.83 $\pm$ 3.91 & \\
\texttt{$\circ$} & 5844.99 $\pm$ 0.13 & 5844.96 (0.63)$^{\mathrm{a}}$, 5844.80$^{\mathrm{b}}$, 5844.95 (0.65)$^{\mathrm{c}}$, 5844.80$^{\mathrm{d}}$, 5844.89 (0.46)$^{\mathrm{f}}$ & 0.64 $\pm$ 0.13 & & 3.99 $\pm$ 0.34 & 1.26 $\pm$ 0.15 & $\dagger$\\
\texttt{+} & 5849.91 $\pm$ 0.08 & 5849.88 (0.93)$^{\mathrm{a}}$, 5849.80$^{\mathrm{b}}$, 5849.81 (1.07)$^{\mathrm{c}}$, 5849.85$^{\mathrm{d}}$, 5849.78 (1.03)$^{\mathrm{e}}$, 5849.82 (0.83)$^{\mathrm{f}}$ & 0.97 $\pm$ 0.15 & & 42.93 $\pm$ 1.35 & 1.88 $\pm$ 0.58 & \\
\texttt{$\circ$} & 5854.11 $\pm$ 1.24 & 5854.54 (0.45)$^{\mathrm{a}}$, 5854.50$^{\mathrm{b}}$, 5854.85 (2.13)$^{\mathrm{c}}$, 5854.54$^{\mathrm{d}}$, 5853.95 (1.79)$^{\mathrm{e}}$, 5854.58 (0.55)$^{\mathrm{f}}$ & 3.04 $\pm$ 2.39 & & 7.55 $\pm$ 1.95 & 3.08 $\pm$ 0.85 & \\
\texttt{-} & 5857.77 $\pm$ 0.78 & 5857.19$^{\mathrm{d}}$ & 2.15 $\pm$ 1.51 & & 3.32 $\pm$ 3.53 & 4.20 $\pm$ 1.53 & \\
\texttt{$\circ$} & 5859.38 $\pm$ 1.11 & 5859.00$^{\mathrm{d}}$, 5859.06 (0.55)$^{\mathrm{f}}$ & 2.68 $\pm$ 2.14 & & 9.85 $\pm$ 1.30 & 2.02 $\pm$ 0.57 & \\
\texttt{+} & 5862.23 $\pm$ 0.75 & 5862.26 (0.50)$^{\mathrm{a}}$, 5861.89$^{\mathrm{d}}$, 5862.21 (0.53)$^{\mathrm{f}}$ & 2.50 $\pm$ 1.47 & & 13.43 $\pm$ 1.89 & 2.51 $\pm$ 0.82 & \\
\texttt{+} & 5864.43 $\pm$ 0.77 & & 2.89 $\pm$ 1.50 & & 20.86 $\pm$ 2.27 & 1.54 $\pm$ 0.98 & $\ddag$\\
\texttt{$\circ$} & 5866.56 $\pm$ 0.94 & 5866.36 (0.70)$^{\mathrm{a}}$, 5866.32 (0.60)$^{\mathrm{f}}$ & 2.65 $\pm$ 1.82 & & 8.28 $\pm$ 2.02 & 3.37 $\pm$ 0.87 & \\
\texttt{+} & 5868.36 $\pm$ 0.78 & & 2.16 $\pm$ 1.49 & & 18.58 $\pm$ 2.66 & -0.59 $\pm$ 1.15 & $\ddag$\\
\texttt{-} & 6483.49 $\pm$ 0.31 & & 2.03 $\pm$ 0.59 & & 11.03 $\pm$ 7.53 & 4.65 $\pm$ 3.55 & \\
\texttt{+} & 6489.59 $\pm$ 0.31 & 6489.38$^{\mathrm{b}}$, 6489.24 (0.64)$^{\mathrm{c}}$, 6489.62$^{\mathrm{d}}$, 6489.42 (0.89)$^{\mathrm{f}}$ & 1.42 $\pm$ 0.59 & & 9.39 $\pm$ 2.03 & 1.59 $\pm$ 0.96 & \\
\texttt{+} & 6492.03 $\pm$ 0.12 & 6492.14 (0.74)$^{\mathrm{a}}$, 6492.02$^{\mathrm{b}}$, 6492.17 (0.78)$^{\mathrm{c}}$, 6492.15$^{\mathrm{d}}$, 6491.88 (0.76)$^{\mathrm{e}}$, 6492.01 (0.57)$^{\mathrm{f}}$ & 0.39 $\pm$ 0.22 & & 5.18 $\pm$ 0.25 & 0.02 $\pm$ 0.12 & \\
\texttt{-} & 6493.11 $\pm$ 0.26 & 6494.14 (9.20)$^{\mathrm{a}}$, 6492.92 (9.67)$^{\mathrm{c}}$, 6494.17 (11.17)$^{\mathrm{e}}$, 6494.13 (7.40)$^{\mathrm{f}}$, 6494 (11.17)$^{\mathrm{g}}$ & 4.33 $\pm$ 0.50 & & 42.82 $\pm$ 13.43 & 16.51 $\pm$ 6.33 & \\
\texttt{+} & 6496.67 $\pm$ 0.20 & 6498.01 (0.79)$^{\mathrm{a}}$, 6497.79 (0.73)$^{\mathrm{c}}$, 6497.82$^{\mathrm{d}}$, 6497.55 (0.50)$^{\mathrm{e}}$, 6498.00 (0.66)$^{\mathrm{f}}$ & 2.77 $\pm$ 0.38 & & 52.89 $\pm$ 9.39 & 2.64 $\pm$ 4.43 & \\
\texttt{$\circ$} & 6501.49 $\pm$ 0.78 & 6501.23$^{\mathrm{d}}$ & 2.58 $\pm$ 1.51 & & 8.13 $\pm$ 1.86 & 1.83 $\pm$ 0.88 & \\
\texttt{+} & 6507.63 $\pm$ 1.45 & & 6.59 $\pm$ 2.78 & & 13.99 $\pm$ 2.72 & 5.52 $\pm$ 1.28 & $\ddag$\\
\texttt{+} & 6516.33 $\pm$ 0.54 & 6516.64$^{\mathrm{d}}$ & 3.72 $\pm$ 0.61 & 0.73 $\pm$ 0.33 & 37.81 $\pm$ 6.54 & 12.23 $\pm$ 3.08 & \\
\texttt{+} & 6518.65 $\pm$ 0.13 & 6518.76$^{\mathrm{d}}$ & 1.41 $\pm$ 0.13 & & 9.66 $\pm$ 0.89 & 4.19 $\pm$ 0.42 & $\dagger$\\
\hline
\end{tabular}
\begin{tablenotes}
      \small
      \item Cross-reference with a DIB noted in \texttt{a}: \citetalias{HobbsYork2009}; \texttt{b}: \citetalias{GalazutdinovMusaev2000}; \texttt{c}: \citetalias{Tuairisg2000}; \texttt{d}: \citetalias{Weselak2000}; \texttt{e}: \citetalias{Jenniskens1994}; \texttt{f}: \citetalias{Fan2019}; \texttt{g}: \citetalias{Sonnentrucker2018}.
    \end{tablenotes}
\end{threeparttable}
\end{table}
\end{landscape}

\begin{landscape}
\begin{table}
\centering
\caption{continued.}
\begin{threeparttable}
\label{tab:table_dibs_4}
\begin{tabular}{l c c c c c c l}
\hline
Q & $\lambda_{C}$ & $\lambda_{\mathrm{REF}}$ ($FWHM_{\mathrm{REF}}$) & $FWHM$ & $\gamma$ & $\Delta EW$ / $\Delta E(\mathrm{B-V})$ & $EW(E\mathrm{(B-V) = 0})$ & Comment\\
& (\AA) & (\AA) & (\AA) & & ($\mathrm{m}$\AA\ / $\mathrm{mag}$) & ($\mathrm{m}$\AA) & \\\hline\hline
\texttt{+} & 6520.67 $\pm$ 0.04 & 6520.75 (1.02)$^{\mathrm{a}}$, 6520.56$^{\mathrm{b}}$, 6520.95 (1.35)$^{\mathrm{c}}$, 6520.70 (0.97)$^{\mathrm{e}}$, 6520.74 (1.00)$^{\mathrm{f}}$ & 0.81 $\pm$ 0.04 & & 13.24 $\pm$ 0.14 & 5.05 $\pm$ 0.06 & $\dagger$\\
\texttt{+} & 6523.32 $\pm$ 0.10 & 6523.36$^{\mathrm{d}}$, 6523.29 (0.48)$^{\mathrm{f}}$ & 1.01 $\pm$ 0.10 & & 6.09 $\pm$ 1.18 & 4.29 $\pm$ 0.56 & $\dagger$\\
\texttt{+} & 6526.11 $\pm$ 0.01 & & 3.17 $\pm$ 0.47 & & 17.49 $\pm$ 1.81 & 13.69 $\pm$ 0.86 & $\ddag$\\
\texttt{+} & 6530.17 $\pm$ 0.24 & 6532.10 (17.20)$^{\mathrm{e}}$, 6532 (17.20)$^{\mathrm{g}}$ & 4.73 $\pm$ 0.44 & -0.10 $\pm$ 0.10 & 51.42 $\pm$ 1.99 & 21.57 $\pm$ 0.94 & \\
\texttt{+} & 6535.89 $\pm$ 0.09 & 6536.51 (1.35)$^{\mathrm{a}}$, 6536.86 (0.69)$^{\mathrm{c}}$, 6536.89$^{\mathrm{d}}$, 6536.44 (0.69)$^{\mathrm{e}}$, 6536.51 (0.86)$^{\mathrm{f}}$ & 5.01 $\pm$ 0.52 & 0.10 $\pm$ 0.10 & 38.57 $\pm$ 1.99 & 23.19 $\pm$ 0.94 & \\
\texttt{+} & 6541.43 $\pm$ 0.12 & & 0.94 $\pm$ 0.12 & & 6.67 $\pm$ 0.50 & 1.63 $\pm$ 0.23 & $\dagger$, $\ddag$\\
\texttt{+} & 6543.00 $\pm$ 0.06 & 6543.29 (0.83)$^{\mathrm{a}}$, 6543.20$^{\mathrm{b}}$, 6543.22 (0.69)$^{\mathrm{f}}$ & 0.56 $\pm$ 0.06 & & 6.62 $\pm$ 0.62 & 0.59 $\pm$ 0.29 & $\dagger$\\
\texttt{-} & 6543.72 $\pm$ 0.17 & 6543.22 (0.69)$^{\mathrm{f}}$ & 0.58 $\pm$ 0.17 & & 0.41 $\pm$ 0.68 & 2.36 $\pm$ 0.32 & $\dagger$\\
\texttt{$\circ$} & 6551.82 $\pm$ 0.12 & & 0.62 $\pm$ 0.12 & & 4.44 $\pm$ 0.85 & -0.18 $\pm$ 0.40 & $\dagger$, $\ddag$\\
\texttt{-} & 6552.47 $\pm$ 0.17 & & 0.35 $\pm$ 0.17 & & 0.66 $\pm$ 0.48 & 0.64 $\pm$ 0.23 & $\dagger$\\
\texttt{+} & 6553.93 $\pm$ 0.13 & 6553.95 (0.57)$^{\mathrm{a}}$, 6553.82$^{\mathrm{b}}$, 6553.76 (0.79)$^{\mathrm{c}}$, 6553.89$^{\mathrm{d}}$, 6553.88 (0.51)$^{\mathrm{f}}$ & 0.49 $\pm$ 0.24 & & 7.04 $\pm$ 1.04 & 0.04 $\pm$ 0.49 & \\
\texttt{-} & 6556.26 $\pm$ 0.21 & & 0.57 $\pm$ 0.21 & & 1.03 $\pm$ 0.79 & 1.13 $\pm$ 0.37 & $\dagger$\\
\texttt{+} & 6567.90 $\pm$ 0.13 & 6567.63$^{\mathrm{b}}$ & 1.26 $\pm$ 0.13 & & 8.09 $\pm$ 0.75 & 3.77 $\pm$ 0.35 & $\dagger$\\
\texttt{+} & 6570.83 $\pm$ 0.57 & 6570.52$^{\mathrm{b}}$ & 3.68 $\pm$ 1.13 & & 30.82 $\pm$ 5.73 & 0.42 $\pm$ 2.70 & \\
\texttt{+} & 6574.08 $\pm$ 0.07 & & 0.46 $\pm$ 0.07 & & 5.91 $\pm$ 0.77 & -0.86 $\pm$ 0.36 & $\dagger$, $\ddag$\\
\texttt{$\circ$} & 6575.78 $\pm$ 0.66 & & 1.40 $\pm$ 1.28 & & 4.62 $\pm$ 1.31 & 0.66 $\pm$ 0.62 & $\ddag$\\
\texttt{$\circ$} & 6578.43 $\pm$ 0.46 & & 0.79 $\pm$ 0.88 & & 4.01 $\pm$ 0.69 & -0.01 $\pm$ 0.32 & $\ddag$\\
\texttt{-} & 6580.46 $\pm$ 0.01 & & 2.67 $\pm$ 1.50 & & 11.33 $\pm$ 5.33 & 1.09 $\pm$ 2.51 & \\
\texttt{+} & 6586.28 $\pm$ 0.39 & & 2.94 $\pm$ 0.67 & & 20.25 $\pm$ 5.91 & 6.04 $\pm$ 2.79 & $\ddag$\\
\texttt{+} & 6589.97 $\pm$ 0.01 & 6590.58 (4.41)$^{\mathrm{a\ast}}$, 6591.03 (4.19)$^{\mathrm{c}}$, 6591.40 (5.55)$^{\mathrm{e}}$ & 4.55 $\pm$ 0.64 & & 50.34 $\pm$ 7.17 & 8.75 $\pm$ 3.38 & \\
\texttt{+} & 6592.53 $\pm$ 0.58 & 6591.51 (7.85)$^{\mathrm{f}}$ & 2.03 $\pm$ 1.11 & & 27.82 $\pm$ 5.62 & -5.41 $\pm$ 2.65 & \\
\texttt{+} & 6594.49 $\pm$ 0.10 & 6594.36 (0.79)$^{\mathrm{a}}$, 6594.13$^{\mathrm{b}}$, 6594.30 (0.71)$^{\mathrm{f}}$ & 1.57 $\pm$ 0.10 & & 17.58 $\pm$ 1.22 & 2.82 $\pm$ 0.58 & $\dagger$\\
\texttt{+} & 6597.37 $\pm$ 0.09 & 6597.43 (0.69)$^{\mathrm{a}}$, 6597.31$^{\mathrm{b}}$, 6597.47 (0.79)$^{\mathrm{c}}$, 6597.34$^{\mathrm{d}}$, 6597.39 (0.72)$^{\mathrm{e}}$, 6597.34 (0.60)$^{\mathrm{f}}$ & 0.77 $\pm$ 0.18 & & 3.20 $\pm$ 0.22 & 5.70 $\pm$ 0.39 & \\
\texttt{$\circ$} & 6598.57 $\pm$ 0.13 & & 0.79 $\pm$ 0.13 & & 5.83 $\pm$ 0.62 & 0.07 $\pm$ 0.29 & $\dagger$, $\ddag$\\
\texttt{$\circ$} & 6600.28 $\pm$ 0.60 & 6600.09 (0.65)$^{\mathrm{a}}$, 6599.99 (0.50)$^{\mathrm{f}}$ & 1.80 $\pm$ 2.24 & -3.00 $\pm$ 9.75 & 9.60 $\pm$ 2.59 & -0.82 $\pm$ 1.22 & \\
\texttt{+} & 6604.11 $\pm$ 0.89 & & 3.31 $\pm$ 1.71 & & 16.16 $\pm$ 2.00 & 0.20 $\pm$ 0.94 & $\ddag$\\
\texttt{$\circ$} & 6606.85 $\pm$ 0.23 & 6607.09 (0.60)$^{\mathrm{a}}$, 6607.10 (0.53)$^{\mathrm{f}}$ & 1.27 $\pm$ 0.23 & & 5.61 $\pm$ 1.83 & 1.10 $\pm$ 0.86 & $\dagger$\\
\texttt{$\circ$} & 6611.32 $\pm$ 0.32 & 6611.06 (1.06)$^{\mathrm{a}}$, 6611.14 (1.39)$^{\mathrm{f}}$ & 1.41 $\pm$ 0.62 & & 6.90 $\pm$ 1.81 & 0.72 $\pm$ 0.85 & \\
\texttt{+} & 6613.66 $\pm$ 0.01 & 6613.70 (1.08)$^{\mathrm{a}}$, 6613.56$^{\mathrm{b}}$, 6613.63 (1.14)$^{\mathrm{c}}$, 6613.52$^{\mathrm{d}}$, 6613.72 (1.14)$^{\mathrm{e}}$, 6613.74 (1.05)$^{\mathrm{f}}$ & 1.02 $\pm$ 0.02 & -0.57 $\pm$ 0.08 & 130.37 $\pm$ 4.38 & 31.90 $\pm$ 2.06 & \\
\texttt{-} & 6616.34 $\pm$ 0.67 & 6616.12 (0.61)$^{\mathrm{f}}$ & 3.71 $\pm$ 1.20 & -0.03 $\pm$ 0.42 & 1.64 $\pm$ 4.72 & 6.47 $\pm$ 2.23 & \\
\texttt{$\circ$} & 6618.49 $\pm$ 0.22 & 6618.38 (0.84)$^{\mathrm{a\ast}}$ & 1.12 $\pm$ 0.22 & & 4.74 $\pm$ 0.67 & 0.87 $\pm$ 0.32 & $\dagger$\\
\texttt{$\circ$} & 6621.56 $\pm$ 0.43 & 6621.71 (0.67)$^{\mathrm{a}}$, 6621.69 (0.54)$^{\mathrm{f}}$ & 1.38 $\pm$ 1.18 & 2.05 $\pm$ 4.79 & 5.11 $\pm$ 0.58 & 1.15 $\pm$ 0.27 & \\
\texttt{+} & 6622.89 $\pm$ 0.15 & 6622.92 (0.71)$^{\mathrm{a}}$, 6622.59 (2.07)$^{\mathrm{c}}$, 6622.90$^{\mathrm{d}}$, 6622.84 (0.58)$^{\mathrm{f}}$ & 0.77 $\pm$ 0.29 & & 5.52 $\pm$ 0.32 & 2.26 $\pm$ 0.15 & \\
\texttt{+} & 6625.54 $\pm$ 0.53 & 6625.79 (0.67)$^{\mathrm{a}}$, 6624.90 (0.70)$^{\mathrm{f}}$, 6625.82 (0.63)$^{\mathrm{f}}$ & 2.18 $\pm$ 1.01 & & 10.78 $\pm$ 1.47 & 1.86 $\pm$ 0.69 & \\
\texttt{$\circ$} & 6628.24 $\pm$ 0.56 & 6627.88$^{\mathrm{d}}$, 6628.24 (0.99)$^{\mathrm{f}}$ & 1.38 $\pm$ 1.08 & & 4.05 $\pm$ 0.74 & 1.26 $\pm$ 0.35 & \\
\texttt{-} & 6630.83 $\pm$ 0.34 & 6630.82 (0.56)$^{\mathrm{a}}$, 6630.80$^{\mathrm{b}}$, 6630.81 (0.55)$^{\mathrm{f}}$ & 0.51 $\pm$ 0.65 & & 4.89 $\pm$ 1.63 & -0.73 $\pm$ 0.77 & \\
\texttt{-} & 6631.64 $\pm$ 0.67 & 6631.71 (0.48)$^{\mathrm{a}}$, 6631.66$^{\mathrm{b}}$, 6631.65 (0.48)$^{\mathrm{f}}$ & 0.38 $\pm$ 1.26 & & 2.83 $\pm$ 0.20 & -0.67 $\pm$ 0.09 & \\
\texttt{$\circ$} & 6632.84 $\pm$ 0.14 & 6632.93 (1.13)$^{\mathrm{a}}$, 6632.85$^{\mathrm{b}}$, 6632.86 (0.95)$^{\mathrm{c}}$, 6632.65$^{\mathrm{d}}$, 6632.93 (1.06)$^{\mathrm{e}}$, 6633.00 (1.08)$^{\mathrm{f}}$ & 0.97 $\pm$ 0.14 & & 3.83 $\pm$ 0.51 & 3.22 $\pm$ 0.24 & $\dagger$\\
\texttt{-} & 6637.76 $\pm$ 0.83 & 6637.82 (0.63)$^{\mathrm{a}}$, 6637.74 (0.80)$^{\mathrm{f}}$ & 0.49 $\pm$ 1.57 & & -0.40 $\pm$ 4.62 & 0.56 $\pm$ 2.18 & \\
\texttt{+} & 6643.20 $\pm$ 0.61 & 6643.64 (1.76)$^{\mathrm{a}}$, 6643.43 (1.74)$^{\mathrm{f}}$ & 4.36 $\pm$ 1.79 & 0.83 $\pm$ 0.84 & 19.33 $\pm$ 3.88 & 4.68 $\pm$ 1.83 & \\
\texttt{+} & 6645.74 $\pm$ 0.20 & 6646.05 (1.16)$^{\mathrm{a}}$, 6646.03$^{\mathrm{b}}$, 6645.95 (0.58)$^{\mathrm{c}}$, 6645.99 (0.84)$^{\mathrm{f}}$ & 1.43 $\pm$ 0.20 & & 7.60 $\pm$ 1.18 & 1.59 $\pm$ 0.55 & $\dagger$\\
\hline
\end{tabular}
\begin{tablenotes}
      \small
      \item Cross-reference with a DIB noted in \texttt{a}: \citetalias{HobbsYork2009}; \texttt{b}: \citetalias{GalazutdinovMusaev2000}; \texttt{c}: \citetalias{Tuairisg2000}; \texttt{d}: \citetalias{Weselak2000}; \texttt{e}: \citetalias{Jenniskens1994}; \texttt{f}: \citetalias{Fan2019}; \texttt{a$\ast$}: Cross-reference with a DIB noted in \citetalias{HobbsYork2009} (Table 3) defined as a possible DIB.
    \end{tablenotes}
\end{threeparttable}
\end{table}
\end{landscape}

\begin{landscape}
\begin{table}
\centering
\caption{continued.}
\begin{threeparttable}
\label{tab:table_dibs_5}
\begin{tabular}{l c c c c c c l}
\hline
Q & $\lambda_{C}$ & $\lambda_{\mathrm{REF}}$ ($FWHM_{\mathrm{REF}}$) & $FWHM$ & $\gamma$ & $\Delta EW$ / $\Delta E(\mathrm{B-V})$ & $EW(E\mathrm{(B-V) = 0})$ & Comment\\
& (\AA) & (\AA) & (\AA) & & ($\mathrm{m}$\AA\ / $\mathrm{mag}$) & ($\mathrm{m}$\AA) & \\\hline\hline
\texttt{$\circ$} & 6649.88 $\pm$ 0.26 & 6649.90 (1.39)$^{\mathrm{f}}$ & 1.39 $\pm$ 0.26 & & 6.34 $\pm$ 1.09 & 0.22 $\pm$ 0.51 & $\dagger$\\
\texttt{$\circ$} & 6651.68 $\pm$ 0.19 & 6651.90 (0.69)$^{\mathrm{a\ast}}$ & 1.16 $\pm$ 0.19 & & 5.91 $\pm$ 0.84 & 1.00 $\pm$ 0.40 & $\dagger$\\
\texttt{$\circ$} & 6654.54 $\pm$ 0.43 & 6654.63 (1.12)$^{\mathrm{a}}$, 6654.58$^{\mathrm{b}}$, 6654.72 (0.70)$^{\mathrm{f}}$ & 1.63 $\pm$ 0.82 & & 4.78 $\pm$ 1.35 & 2.86 $\pm$ 0.64 & \\
\texttt{+} & 6657.32 $\pm$ 0.18 & 6657.47 (1.08)$^{\mathrm{a}}$, 6657.34 (0.80)$^{\mathrm{f}}$ & 1.23 $\pm$ 0.18 & & 6.25 $\pm$ 0.77 & 1.92 $\pm$ 0.36 & $\dagger$\\
\texttt{$\circ$} & 6658.75 $\pm$ 0.38 & 6658.77 (0.67)$^{\mathrm{a}}$, 6658.70 (0.72)$^{\mathrm{f}}$ & 1.62 $\pm$ 0.73 & & 5.68 $\pm$ 0.72 & 3.01 $\pm$ 0.34 & \\
\texttt{+} & 6660.74 $\pm$ 0.04 & 6660.73 (0.67)$^{\mathrm{a}}$, 6660.64$^{\mathrm{b}}$, 6660.62 (0.82)$^{\mathrm{c}}$, 6660.70$^{\mathrm{d}}$, 6660.64 (0.84)$^{\mathrm{e}}$, 6660.67 (0.63)$^{\mathrm{f}}$ & 0.66 $\pm$ 0.07 & 0.36 $\pm$ 0.76 & 18.61 $\pm$ 2.08 & 7.29 $\pm$ 0.98 & \\
\texttt{$\circ$} & 6662.18 $\pm$ 0.21 & 6662.29 (0.65)$^{\mathrm{a}}$, 6662.25 (0.43)$^{\mathrm{c}}$, 6662.17 (0.46)$^{\mathrm{f}}$ & 1.25 $\pm$ 0.21 & & 3.81 $\pm$ 1.13 & 3.40 $\pm$ 0.53 & $\dagger$\\
\texttt{+} & 6665.18 $\pm$ 0.11 & 6665.25 (0.57)$^{\mathrm{a\ast}}$, 6665.15$^{\mathrm{b}}$, 6665.15$^{\mathrm{d}}$, 6665.27 (0.72)$^{\mathrm{f}}$ & 0.79 $\pm$ 0.11 & & 5.26 $\pm$ 0.12 & 1.42 $\pm$ 0.06 & $\dagger$\\
\texttt{+} & 6672.28 $\pm$ 0.11 & 6672.30 (0.96)$^{\mathrm{a\ast}}$, 6672.15$^{\mathrm{b}}$, 6672.39$^{\mathrm{d}}$, 6672.23 (0.67)$^{\mathrm{f}}$ & 0.83 $\pm$ 0.21 & & 10.74 $\pm$ 0.50 & 2.68 $\pm$ 0.24 & \\
\texttt{-} & 6674.06 $\pm$ 0.72 & & 1.35 $\pm$ 1.39 & & 1.70 $\pm$ 1.12 & 1.48 $\pm$ 0.53 & \\
\texttt{-} & 6680.74 $\pm$ 0.45 & 6681.07 (2.01)$^{\mathrm{c}}$ & 2.67 $\pm$ 1.05 & & 5.36 $\pm$ 3.68 & 5.88 $\pm$ 1.74 & \\
\texttt{+} & 6685.38 $\pm$ 1.00 & 6684.91 (1.64)$^{\mathrm{a}}$, 6684.83 (1.09)$^{\mathrm{c}}$, 6685.03 (1.42)$^{\mathrm{f}}$ & 2.57 $\pm$ 1.93 & & 12.37 $\pm$ 3.02 & -0.56 $\pm$ 1.42 & \\
\texttt{+} & 6689.37 $\pm$ 0.28 & 6689.38 (0.99)$^{\mathrm{a}}$, 6689.30$^{\mathrm{b}}$, 6689.38 (0.61)$^{\mathrm{c}}$, 6689.35 (0.89)$^{\mathrm{f}}$ & 1.15 $\pm$ 0.54 & & 9.22 $\pm$ 0.95 & 0.88 $\pm$ 0.45 & \\
\texttt{$\circ$} & 6693.55 $\pm$ 0.21 & 6693.63 (0.73)$^{\mathrm{a}}$, 6693.35$^{\mathrm{b}}$, 6693.57 (0.55)$^{\mathrm{f}}$ & 1.34 $\pm$ 0.21 & & 6.66 $\pm$ 0.73 & 1.08 $\pm$ 0.34 & $\dagger$\\
\texttt{$\circ$} & 6694.47 $\pm$ 0.09 & 6694.60 (0.65)$^{\mathrm{a}}$, 6694.48$^{\mathrm{b}}$, 6694.40 (1.19)$^{\mathrm{c}}$, 6694.50$^{\mathrm{d}}$, 6694.46 (0.64)$^{\mathrm{e}}$, 6694.53 (0.62)$^{\mathrm{f}}$ & 0.57 $\pm$ 0.09 & & 4.76 $\pm$ 0.16 & 0.55 $\pm$ 0.07 & $\dagger$\\
\texttt{+} & 6699.33 $\pm$ 0.09 & 6699.36 (0.82)$^{\mathrm{a}}$, 6699.24$^{\mathrm{b}}$, 6699.28 (0.99)$^{\mathrm{c}}$, 6699.24$^{\mathrm{d}}$, 6699.37 (1.15)$^{\mathrm{e}}$, 6699.28 (0.67)$^{\mathrm{f}}$ & 0.99 $\pm$ 0.16 & 0.69 $\pm$ 0.84 & 25.15 $\pm$ 6.45 & 2.61 $\pm$ 0.67 & \\
\texttt{+} & 6701.97 $\pm$ 0.07 & 6702.15 (0.74)$^{\mathrm{a}}$, 6701.98$^{\mathrm{b}}$, 6701.87 (0.83)$^{\mathrm{c}}$, 6701.95$^{\mathrm{d}}$, 6701.98 (0.99)$^{\mathrm{e}}$, 6702.07 (0.74)$^{\mathrm{f}}$ & 0.87 $\pm$ 0.07 & & 10.93 $\pm$ 0.23 & 1.14 $\pm$ 0.11 & $\dagger$\\
\texttt{-} & 6706.93 $\pm$ 0.70 & 6706.61 (0.96)$^{\mathrm{a}}$, 6706.48 (0.91)$^{\mathrm{f}}$ & 1.53 $\pm$ 1.36 & & 2.26 $\pm$ 3.27 & 1.78 $\pm$ 1.54 & \\
\texttt{+} & 6707.93 $\pm$ 0.18 & 6707.73 (1.38)$^{\mathrm{c}}$, 6707.90$^{\mathrm{d}}$ & 0.67 $\pm$ 0.34 & & 6.40 $\pm$ 1.89 & 0.69 $\pm$ 0.89 & Li I\\
\texttt{-} & 6709.59 $\pm$ 0.36 & 6709.54 (0.96)$^{\mathrm{a}}$, 6709.39$^{\mathrm{b}}$, 6709.65 (1.76)$^{\mathrm{c}}$, 6709.44$^{\mathrm{d}}$, 6709.24 (1.30)$^{\mathrm{e}}$, 6709.49 (0.71)$^{\mathrm{f}}$ & 1.42 $\pm$ 0.68 & & 7.86 $\pm$ 3.40 & 1.78 $\pm$ 1.60 & \\
\texttt{+} & 6713.88 $\pm$ 0.79 & 6713.81 (1.13)$^{\mathrm{a}}$, 6713.79 (1.00)$^{\mathrm{f}}$ & 3.18 $\pm$ 1.52 & & 22.58 $\pm$ 1.01 & -1.19 $\pm$ 0.48 & \\
\texttt{+} & 6719.67 $\pm$ 0.74 & 6719.25 (0.71)$^{\mathrm{a}}$, 6719.58 (3.05)$^{\mathrm{c}}$, 6719.19 (0.51)$^{\mathrm{f}}$ & 4.12 $\pm$ 1.46 & & 21.23 $\pm$ 1.79 & 3.12 $\pm$ 0.85 & \\
\texttt{+} & 6724.45 $\pm$ 0.87 & 6724.14 (0.67)$^{\mathrm{a\ast}}$ & 4.32 $\pm$ 1.52 & & 23.75 $\pm$ 0.59 & 1.54 $\pm$ 0.28 & \\
\texttt{+} & 6727.95 $\pm$ 0.45 & 6727.68 (0.71)$^{\mathrm{a}}$, 6727.64 (0.73)$^{\mathrm{f}}$ & 1.97 $\pm$ 0.89 & & 7.52 $\pm$ 0.89 & 3.11 $\pm$ 0.42 & \\
\texttt{+} & 6729.32 $\pm$ 0.24 & 6729.30 (0.67)$^{\mathrm{a}}$, 6729.28$^{\mathrm{b}}$, 6729.20 (0.81)$^{\mathrm{c}}$, 6729.28$^{\mathrm{d}}$, 6729.22 (0.60)$^{\mathrm{f}}$ & 0.75 $\pm$ 0.47 & & 8.23 $\pm$ 1.12 & -0.41 $\pm$ 0.53 & \\
\texttt{$\circ$} & 6731.30 $\pm$ 0.52 & 6731.21 (0.92)$^{\mathrm{a}}$, 6731.28 (1.08)$^{\mathrm{f}}$ & 1.45 $\pm$ 1.00 & & 3.97 $\pm$ 1.05 & 1.88 $\pm$ 0.50 & \\
\texttt{+} & 6732.98 $\pm$ 0.41 & 6733.25 (1.27)$^{\mathrm{f}}$ & 1.62 $\pm$ 0.80 & & 11.13 $\pm$ 0.88 & 0.92 $\pm$ 0.41 & \\
\texttt{-} & 6733.70 $\pm$ 0.60 & 6733.38 (1.24)$^{\mathrm{a}}$, 6733.35 (1.28)$^{\mathrm{c}}$, 6733.25 (1.27)$^{\mathrm{f}}$ & 1.40 $\pm$ 1.15 & & 10.44 $\pm$ 3.31 & -1.05 $\pm$ 1.56 & \\
\hline
\end{tabular}
\begin{tablenotes}
      \small
      \item Cross-reference with a DIB noted in \texttt{a}: \citetalias{HobbsYork2009}; \texttt{b}: \citetalias{GalazutdinovMusaev2000}; \texttt{c}: \citetalias{Tuairisg2000}; \texttt{d}: \citetalias{Weselak2000}; \texttt{e}: \citetalias{Jenniskens1994}; \texttt{f}: \citetalias{Fan2019};
      \texttt{a$\ast$}: Cross-reference with a DIB noted in \citetalias{HobbsYork2009} (Table 3) defined as a possible DIB.
    \end{tablenotes}
\end{threeparttable}
\end{table}
\end{landscape}

\bsp	
\label{lastpage}
\end{document}